\documentclass[aps,prd,reprint,showkeys,nofootinbib,superscriptaddress,notitlepage]{revtex4-1}
\usepackage{amsmath,graphicx,amssymb}
\usepackage[utf8]{inputenc}
\usepackage[unicode=true,
  bookmarks=false,   backref=false,    colorlinks=true,
  linktocpage=true,  citecolor=black,  linkcolor=black,
  urlcolor=black,    breaklinks=true
]{hyperref}

\newcommand{\GN}{G_{\rm N}}
\newcommand{\relG}{\Delta G/\GN}
\newcommand{\lC}{\lambda_{\rm C}}
\newcommand{\lChat}{\hat{\lambda}_\mathrm{C}}
\newcommand{\rhat}{\hat{r}}

\newcommand{\mpl}{M_\mathrm{Pl}}

\newcommand{\relR}{\frac{\Delta R}{R}}
\newcommand{\sa}{s_\alpha}
\newcommand{\FN}{F_\mathrm{N}}
\newcommand{\zmin}{z_\mathrm{min}}
\newcommand{\zmax}{z_\mathrm{max}}
\newcommand{\zlos}{\hat{z}_\mathrm{LOS}}

\newcommand{\be}{\begin{equation}}
\newcommand{\ee}{\end{equation}}

\begin{document}

\title{Impact of symmetron screening on the Hubble tension: new constraints using cosmic distance ladder data}

\author{Marcus \surname{Högås}}
\email{marcus.hogas@fysik.su.se}
\affiliation{Oskar Klein Centre, Department of Physics, Stockholm University\\Albanova University Center\\ 106 91 Stockholm, Sweden}
\affiliation{Department of Mathematics, Stockholm University\\ 106 91 Stockholm, Sweden}

\author{Edvard \surname{Mörtsell}}
\email{edvard@fysik.su.se}
\affiliation{Oskar Klein Centre, Department of Physics, Stockholm University\\Albanova University Center\\ 106 91 Stockholm, Sweden}

\begin{abstract}
Fifth forces are ubiquitous in modified theories of gravity. To be compatible with observations, such a force must be screened on solar-system scales but may still give a significant contribution on galactic scales. If this is the case, the fifth force can influence the calibration of the cosmic distance ladder, hence changing the inferred value of the Hubble constant $H_0$. In this paper, we analyze symmetron screening and show that it generally increases the Hubble tension. On the other hand, by doing a full statistical analysis, we show that cosmic distance ladder data are able to constrain the theory to a level competitive with solar-system tests -- currently the most constraining tests of the theory. For the standard coupling case, the constraint on the symmetron  Compton wavelength is $\lC \lesssim 2.5 \, \mathrm{Mpc}$. Thus, distance ladder data constitutes a novel and powerful way of testing this, and similar, types of theories.
\end{abstract}

\maketitle

\section{Introduction}
During the last decade, the cosmological standard model has entered a small crisis involving the present expansion rate of the Universe, parameterized by the Hubble constant $H_0$. The SH0ES team derive $H_0 = (73.0 \pm 1.0) \, \mathrm{km/s/Mpc}$ \citep{Riess:2021jrx}, using type Ia supernovae (SNIae) calibrated with Cepheid variable stars and anchor galaxies with direct distance measurements. This is commonly referred to as a local Hubble estimate. On the other hand, the value inferred from observations of the cosmic microwave background (CMB) with the \emph{Planck} satellite, is $H_0 = (67.8 \pm 0.5) \, \mathrm{km/s/Mpc}$ \citep{Planck2020}. These two results are discrepant to a degree of $\sim 5\sigma$, commonly referred to as the Hubble tension.

Potentially, the difference can be partly attributed to systematic uncertainties, see for example refs.~\cite{mortsell2021hubble,Mortsell:2021tcx}, but may signal the need to reevaluate the assumptions of the cosmological standard model. An intriguing possibility is that the discrepancy is due to new physics modifying the expansion history of the Universe, thus affecting the inferred value of $H_0$ from the \emph{Planck} data. A common idea is that this can be achieved via a new theory of gravity. A successful model would push up the ``low'' \emph{Planck} value towards the SH0ES result without violating other cosmological data sets, see ref.~\cite{Abdalla:2022yfr} for some examples. So far, there is no such consensus solution. An alternative way to modify the expansion history is to add a new component to the energy budget of the Universe. Early dark energy may be a promising candidate 
(see e.g. refs.~\cite{Karwal:2016vyq,Poulin:2018dzj,Poulin:2018cxd,Agrawal:2019lmo,Lin:2019qug,Kamionkowski:2022pkx}).

Another possibility is that we are located in an underdense region of the Universe. This is (at least to some degree) supported by data \citep{Keenan:2013mfa,Frith:2003tb,Whitbourn:2013mwa,Bohringer:2019tyj,Wong:2021fvu} and would result in a locally increased expansion rate compared with the cosmological background 
(see e.g. ref.~\cite{Sundell:2015cza}), potentially explaining the 
higher value for $H_0$ obtained from local observations by the SH0ES team.
This is an attractive resolution in the sense that it does not require the introduction of new physics. Unfortunately, when taking the full range of cosmological observations into account, the solution is ruled out \citep{Odderskov:2014hqa,Wu:2017fpr,Kenworthy:2019qwq,Camarena:2021mjr,Castello:2021uad}. 

In this paper, we take a different approach and explore how the SH0ES value for $H_0$ is affected by the presence of a fifth force; an extra degree of freedom modifying the laws of gravity. If the fifth force is stronger in Type Ia supernova (SNIa) host galaxies compared with anchor galaxies with direct distance measurements, the local $H_0$ measurement will be biased to a higher value. Thus, correcting for the fifth force decreases the inferred value of $H_0$, potentially harmonizing it with the \emph{Planck} value, as first suggested in ref.~\cite{Desmond_2019}. Since fifth forces are ubiquitous in modified theories of gravity, this could be an attractive candidate solution to the tension.

In ref.~\cite{Desmond_2019}, the authors find that an additional fifth force strength of 5\%-30\% in the SNIa host galaxies can alleviate the $H_0$ tension. Similar results were found in ref.~\cite{Hogas:2023pjz}, presenting a full statistical analysis. Here, we further develop this idea. First and foremost, in refs.~\cite{Desmond_2019,Hogas:2023pjz}, the strength of the fifth force was parameterized using proxy fields (such as the gravitational potential). Here, we assume a symmetron screening model and explicitly estimate the strength of the fifth force from the theory itself, for each source entering the local distance ladder. Another difference with ref.~\cite{Desmond_2019} is that they estimated the effect of the fifth force on $H_0$ using an effective rescaling of $H_0$. In this paper, we infer the value of $H_0$ using a full statistical data analysis. This also opens up for the possibility of using the distance ladder data to constrain fifth force models observationally, a possibility we capitalize on in this paper.

\section{Executive summary}
The symmetron is a fifth force mediated by a scalar field. The effective potential of the scalar field depends on the ambient energy density, resulting in the force being screened on solar-system scales and unscreened in less dense environments such as in galaxies. In this paper, we (re-)calibrate the Cepheid-based cosmic distance ladder in the presence of a symmetron fifth force, performing a full statistical data analysis. In unscreened (i.e., small mass) galaxies, the fifth force effectively increases the gravitational constant, modifying the Cepheid astrophysics. Rather than approximating the fifth force using some proxy field, we calculate it by solving the equations of motion under the relevant conditions. This results in systematic differences in the period-luminosity relation (PLR) between screened and unscreened galaxies. Taking the fifth force into account ultimately leads to a different inferred value for the Hubble constant, $H_0$. However, it turns out that the value of $H_0$ shifts in the ``wrong'' direction, further exacerbating the Hubble tension rather than solving it.

On the other hand, since we are doing a full statistical data analysis, it results in new observational constraints on the parameter space that the theory must satisfy in order to be compatible with the distance ladder data. Here, we analyze the full three-dimensional parameter space of the symmetron model. 
We also compare the inferred distances to galaxies where there are observations of both Cepheids and tip of the red-giant branch stars (TRGBs). Since the distance modification due to the fifth force acts in opposite directions using Cepheids versus TRGBs, this allows for a consistency tests. The resulting constraints are similar to those from the Cepheid-based distance ladder, see Figs.~\ref{fig:symk2vslambdaAll}-\ref{fig:symk2vslambda_combinedAll}.

For the standard coupling case, we constrain the Compton wavelength of the scalar field to $\lC \lesssim 2.5 \, \mathrm{Mpc}$, comparable to the constraints from solar-system tests.

\section{Theory}
Generically, a modification of GR introduces an additional degree of freedom \citep{Weinberg:1965rz}. In a wide range of modified gravity theories, this new degree of freedom is encoded in a scalar field that couples to gravity. The coupling of the scalar to matter introduces a fifth force which effectively enhances the gravitational force. To be compatible with solar-system tests of gravity, these scalar fields must exhibit a screening mechanism suppressing the fifth force in our (on average, dense) vicinity. However, in less dense environments, fifth forces can be significant, leading to potentially observable effects. 

The symmetron model was first suggested in ref.~\citep{Hinterbichler:2010es} (see \cite{Dehnen1992,Gessner1992} for earlier works on similar models). Currently, the tightest constraints on the symmetron model comes from solar-system tests. Since the surface gravitational potentials of the Sun and the Milky Way (MW) are of the same order of magnitude, the requirement that solar-system tests are satisfied coincides with our galaxy being mostly screened, see ref.~\cite{Hinterbichler:2010es} for details. If the MW is on the verge of being screened, some sources in other galaxies in the distance ladder will be screened and others will not. Thus, the symmetron model can exhibit effects with consequences for the calibration of $H_0$, while still satisfying solar-system tests. This is the reason why we study this model in particular detail. 

In the symmetron model, the fifth force is mediated by a scalar field $\phi$ with a matter coupling that depends on the local matter density; effectively coupling to matter only in low-density environments. In high-density environments, such as inside a star, the coupling to matter is negligible and the Newtonian force of gravity is restored. This feature is realized by postulating the following effective potential for the scalar field,
\begin{equation}
    \label{eq:Veff}
    V_\mathrm{eff}(\phi) = \frac{1}{2} \left( \frac{\rho}{M^2} - \mu^2 \right) \phi^2 + \frac{\lambda}{4} \phi^4,
\end{equation}
where $\rho$ is the matter density (here, we restrict ourselves to non-relativistic matter contents). There are three theory parameters: $M$ and $\mu$ are mass scales whereas $\lambda$ is dimensionless. In low-density regions where $\rho < \mu^2 M^2$, the effective potential takes the form of a ``mexican hat'' with the vacuum expectation value (VEV) of $\phi$ given by,
\begin{equation}
    \label{eq:VEVlow}
    \bar{\phi} = \mu / \sqrt{\lambda} \equiv \phi_0, \quad \rho < \mu^2 M^2,
\end{equation}
thereby breaking the $\mathbb{Z}_2$ symmetry $\phi \to -\phi$. In regions of high matter density where $\rho > \mu^2 M^2$, the minimum of the potential is,
\begin{equation}
    \label{eq:VEVhigh}
    \bar{\phi} = 0, \quad \rho > \mu^2 M^2,
\end{equation}
thereby restoring the $\mathbb{Z}_2$ symmetry in high-density environments (hence, the name ``symmetron'').

The density for which the scalar field becomes tachyonic is $\rho = \mu^2 M^2$, set by the magnitude of the dimensionless parameter,
\begin{equation}
    \sa \equiv \frac{\rho_c}{3M^2 \mu^2}.
\end{equation}
Here, $\rho_c \equiv 3 \mpl^2 H_0^2$ is the present-day critical density of the Universe and $\mpl = (8 \pi G_N)^{-1/2}$ is the Planck mass. 
For the symmetron field to act as dark energy driving the accelerated expansion of the Universe, one should set $\sa \sim 1$, making the scalar field becoming tachyonic for cosmological densities $\rho \sim \rho_c /3$. Since we are interested in the effects of the scalar field on galactic scales, we let $\sa$ be a free parameter. 

The coupling of the symmetron to matter is determined by $\bar{\phi}/M^2$, vanishing in high-density regions where $\bar{\phi} = 0$ in which GR is restored. On the other hand, in low-density environments where $\bar{\phi} = \mu / \sqrt{\lambda}$, a fifth force is mediated. We make use of the dimensionless parameter, $g$, to set the absolute scale for the strength of the fifth force 
\begin{equation}
    g \equiv \frac{\mu \mpl}{\sqrt{\lambda} M^2},
\end{equation}
where $g \sim 1$ corresponds to the fifth force being comparable to the gravitational force in low-density environments \citep[cf.][]{Hinterbichler:2010es}.

Finally, the range of the fifth force in vacuum is set by the Compton wavelength associated with the mass scale of the theory, $\lC \equiv \mu^{-1}$ \citep{Hinterbichler:2010es}. 

To summarize, we have reparameterized the symmetron model from $(M,\mu,\lambda)$ to a set of parameters $(\sa,\lC,g)$ with immediate physical interpretations:
\begin{subequations}
    \begin{align}
    \sa &\equiv \frac{\rho_c}{3M^2 \mu^2},\\
    \lC &\equiv \frac{1}{\mu},\\
    g &\equiv \frac{\mu \mpl}{\sqrt{\lambda} M^2},
\end{align}
\end{subequations}
where $\sa$ sets the density for which the screening mechanism becomes effective, $\lC$ is the Compton wavelength giving the range of the fifth force, and $g$ is the coupling constant setting the absolute scale for the strength of the fifth force relative to the gravitational force.\footnote{Here, we distinguish between the gravitational force and the fifth force. An alternative perspective that we will also take is to view the symmetron field as modifying the strength of the gravitational force.  See Section~\ref{sec:CalcRelG} for the mathematical details.}

\section{Data description}
With the exception of a few additions and modifications to accommodate symmetron model modifications as explained in subsequent sections, we will make use of the same data sets and methods as employed, and described in more detail, in refs.~\cite{mortsell2021hubble, Mortsell:2021tcx}.

To summarize, for anchor galaxies, we use a distance modulus to the Large Magellanic Cloud (LMC) of $\mu_{\rm LMC}=18.477\pm 0.0263$ derived from double eclipsing binaries \citep{Paczynski:1996dj,Pietrzy_ski_2019,Riess_2019}. For the distance to N4258, we use $\mu_{\rm N4258}=29.397\pm 0.032$ derived from mega-maser observations \citep{Reid_2019}. MW Cepheids data, including GAIA parallax measurements, are from Table~1 in ref.~\cite{Riess_2021}.

Data for Cepheids in the LMC, come from Table~2 in ref.~\cite{Riess_2019}, and for Cepheids in M31 and beyond, from Table~4 in ref.~\cite{Riess:2016jrr}. For all galaxies, the Cepheid data includes the sky positions (RA and Dec) necessary for estimating the symmetron screening effect on a source-to-source basis. 
The galaxy velocity dispersions are obtained from the Extragalactic Distance Database (EDD, \cite{Tully_2009}) and listed in Tab.~\ref{tab:sigmalist}.

Type Ia SN peak magnitudes are from Table~5 in ref.~\cite{Riess:2016jrr}. 

TRGB data comes from ref.~\cite{Freedman_2019}. We estimate the average positions of the RGB stars in each host galaxy from the location of the hatched regions of the galaxy fields depicted in Figs.~1-2 in ref.~\cite{Freedman_2019}.

Given the anchor distances and MW Cepheid parallaxes, the Cepheid magnitudes, color excesses, periods and metallicities, together with the SNIa peak magnitudes, we can derive (among other parameters such as the Cepheid absolute magnitude) the SNIa absolute magnitude, $M_{\rm B}$. Since we are focusing on the impact of the symmetron model, we assume a fixed universal color-luminosity relation (corresponding to $R_{\rm H} = 0.386$) when standardizing Cepheid luminosities, similar to the SH0ES team.

The Hubble constant is finally calculated as 
\be 
H_0=10^{M_{\rm B}/5+a_{\rm B}+5} 
\label{eq:h0} 
\ee
where $a_{\rm B}=0.71273 \pm 0.00176$ is the intercept of the SNIa magnitude-redshift relation \citep{Riess:2016jrr}.

\section{Methods I: calculating $\relG$}
\label{sec:CalcRelG}
When the cosmic distance ladder is calibrated, it is assumed  that there is no (unaccounted for) systematic offset between the Cepheid and SNIa magnitudes between anchor, host, and cosmic flow galaxies. However, if there are fifth forces active on galactic scales, this assumption can be violated. A fifth force is typically parameterized as an effective relative shift of Newton's gravitational constant $\GN$ according to,
\begin{equation}
\label{eq:relGdef}
    \frac{F_5}{\FN} = \frac{G - \GN}{\GN} \equiv \frac{\Delta G}{\GN},
\end{equation}
where $F_5$ is the fifth force and $F_\mathrm{N}$ is the Newtonian force (both per unit mass). Eq.~\eqref{eq:relGdef} can be regarded as the definition of the (modified) effective gravitational constant $G$. To calculate $\relG$, two steps must be carried out in the following order: (1) Calculating the scalar field $\phi$ in the galaxy. (2) Calculating $\phi$ and $\relG$ for the stars that we use in the distance ladder. 

Common for these two steps is that the equations of motion (EoM) for the scalar field must be solved. It is convenient to write the EoM in terms of the dimensionless scalar field,
\begin{equation}
    \label{eq:psiDef}
    \psi(r) \equiv \phi(r) / \phi_0,
\end{equation}
and expressing the length scales in units of some scale $R$ (which we will later take to be the radius of the source which we are interested in),
\begin{equation}
     \lChat \equiv \lC / R, \quad \rhat \equiv r / R.
\end{equation}
Assuming spherical symmetry, the EoM for the scalar field reads,
\begin{equation}
    \label{eq:psiEoM}
    \partial_{\rhat}^2 \psi(\rhat) + \frac{2}{\rhat} \partial_{\rhat} \psi(\rhat) = \left( \alpha(\rhat) - \frac{1}{\lChat^2} \right) \psi(\rhat) + \frac{1}{\lChat^2} \psi^3(\rhat),
\end{equation}
where we have defined the rescaled (dimensionless) energy density,
\begin{equation}
\label{eq:alpha}
    \alpha(\rhat) \equiv 3 \, \sa \, \frac{1}{\lChat^2}\, \frac{\rho(\rhat)}{\rho_c}.
\end{equation}
Note that $\sa$ and $\rho$ only appear in the combination $\sa \rho$, so a rescaling of the theory parameter $\sa$ corresponds to an overall rescaling of all densities.

\subsection{Calculating $\phi$ in the galaxies}
As an approximation for the mass density of a galaxy, we assume an isothermal sphere. We truncate the mass density at a radius of $R_{300}$, defined as the radius at which $\rho=300 \rho_c$, implying the following relation
\begin{equation}
    R_{300} \simeq 67 \frac{\sigma}{100 \mathrm{km/s}} \, h^{-1} \, \mathrm{kpc},
\end{equation}
where $\sigma$ is the galaxy velocity dispersion and $h = H_0 / (100 \, \mathrm{km/s/Mpc})$.

If $\lChat$ is small enough for a given density profile $\rho(r)$, the EoM \eqref{eq:psiEoM} can be solved analytically. If not, they are solved numerically, see Appendix~\ref{sec:AppGalaxies} for details. As boundary conditions we impose $\partial_{\rhat} \psi = 0$ at the center ($\rhat=0$) and that $\psi$ should approach some value $\psi_\mathrm{env,gal}$ asymptotically as $\rhat \to \infty$.

Except for the LMC and the MW, we set $\psi_\mathrm{env,gal}=1$ for all galaxies, i.e., we assume that the scalar field reaches its vacuum value asymptotically outside each galaxy. For the LMC, $\psi_\mathrm{env,gal}$ is set by the local value of $\psi$ in the MW at the position of the LMC. Similarly, the MW lies in the proximity of M31. To take this into account, we make a simple estimate, computing the value of the scalar field due to M31 at the position of the MW (assuming vacuum outside M31 as a first approximation).

\subsection{Calculating $\phi$ and $\relG$ for stars}
For simplicity, the stars are modelled as being homogeneous $\rho(r)=\mathrm{const}$. For this case eq.~\eqref{eq:psiEoM} can be solved analytically, as outlined in Appendix~\ref{sec:AppStars}. To fully specify the solution, we need the dimensionless density $\alpha$, the radius of the star, and the environmental value of the scalar field in the vicinity of the star, $\psi_\mathrm{env}$, which is set by the solution for the scalar field of the galaxy at the position of the star.

For Cepheids, we use the relations in ref.~\cite{Gieren_1999} and ref.~\cite{Anderson:2016txx} to estimate their radii and masses from their periods. From this, we can calculate their density using eq.~\eqref{eq:alphaHom}. As outlined in the previous section, $\psi_\mathrm{env}$ can be obtained by evaluating the host galaxy scalar field $\psi$ at the positions of the stars. In Appendix~\ref{sec:PhysDist}, we described how to obtain these positions.

In spherical symmetry, the fifth force (per unit mass) is,
\begin{equation}
    \label{eq:F5}
    F_5 =  g^2 \, \frac{H_0^2 \lC^2}{\sa} \, \frac{1}{R} \psi(\rhat) \partial_{\rhat} \psi(\rhat),
\end{equation}
see for example ref.~\cite{Clampitt:2011mx}. With the solution for $\psi$, we can finally obtain $\relG$.

Since we only know the 2d projected distances of the Cepheids from their galaxy center (except in the MW where we know the physical distances directly via parallax), this is a source of uncertainty in the value of $\relG$. To take this into account in the statistical data analysis, we implement the uncertainty by simulating different realizations of the physical distance of each individual Cepheid using a Monte Carlo method. The details can be found in Appendix~\ref{sec:PhysDist}.

Another source of uncertainty when estimating $\relG$ is the uncertainty in the velocity dispersion of the host galaxy. Similar to the physical distance, this is implemented using a Monte Carlo method. See Appendix~\ref{sec:VelDisp} for details.

\section{Methods II: (re-)calibrating the cosmic distance ladder}
There are three steps in the Cepheid-based distance ladder and in each of these, a fifth force can potentially affect the calibration, ultimately leading to a different value of $H_0$. The effect of a fifth force can be understood by expressing the Hubble constant as,
\begin{align}
    \label{eq:H0effects}
5\log_{10} H_0 = &5\log_{10} r(z)-5\log_{10} D^{\rm anch} \nonumber\\
&+\Delta m_{\rm SN}-\Delta m_{\rm Ceph},
\end{align}
see refs.~\cite{mortsell2021hubble,Mortsell:2021tcx}. Here, $\Delta m_{\rm SN}$ represents a systematic offset in SNIa magnitudes between host galaxies and cosmic flow galaxies and $\Delta m_{\rm Ceph}$ is a systematic offset in Cepheids magnitudes between the host galaxies and anchor galaxies. In the absence of a fifth force, these offsets are assumed to be zero.

From eq.~\eqref{eq:H0effects}, we see that the inferred value of $H_0$ decreases if:
\begin{enumerate}
\item We increase the independent anchor distances, $D^{\rm anch}$.
\item $\Delta m_{\rm Ceph}>0$. This is achieved if the SNIa host galaxies are less screened than the anchor galaxies, 
i.e. if the fifth force is stronger in the host galaxies. 
In this case host Cepheids appear brighter which must be corrected for by raising $m_{\rm Ceph}^{\rm host}$.
\item $\Delta m_{\rm SN}<0$. If SNIa in the Hubble flow are more unscreened than those in Cepheid hosts, they appear brighter and we need to correct for this by raising $m_{\rm SN}^{\rm flow}$.
\end{enumerate}
For a detailed description of the calibration process employed here, see refs.~\cite{mortsell2021hubble,Mortsell:2021tcx}.

\subsection{Anchor distances}
In ref.~\cite{Desmond_2019}, the MW and N4258 are used as anchor galaxies. Here, we include also a distance estimate to the Large Magellanic Cloud (LMC). In the following, we analyze how the geometrical distance anchors are affected by a fifth force.\\

\noindent \textbf{MW.} The distances to the Cepheids in the MW are derived from their observed parallax. Since this is a geometrical measurement, it is independent of $G$.\\ 

\noindent \textbf{N4258.} The distance estimate to N4258 is based on observations of the position, (line of sight) velocity, and (line of sight) acceleration of water masers close to the center of this galaxy. The model prediction of the velocity and acceleration is based on Keplerian motion of the masers (plus relativistic corrections) in which the mass of the the central black hole, $M_\mathrm{BH}$, and the gravitational constant, $G$, always enter in the combination $GM_\mathrm{BH}$, including the relativistic correction due to gravitational redshift which depends on the Schwarzschild radius $2GM_\mathrm{BH}$, see for example ref.~\cite{Humphreys:2013eja}. Hence, $G$ and $M_\mathrm{BH}$ are degenerate and only the combination $GM_\mathrm{BH}$ is observationally constrained; an increase in $G$ is compensated by a decrease in $M_\mathrm{BH}$. Thus, a value of $G$ different from $\GN$ does not influence the estimated distance to N4258. In any case, due to the dense environment around the water masers, we expect the fifth force to be fully screened there.\\

\noindent \textbf{LMC.} The distance to the LMC is estimated based on observations of detached eclipsing binaries (DEBs). The inferred distance to a DEB is based on the orbital velocity and photometric light curve of the system which gives the physical size of the individual stars \citep{Paczynski:1996dj}. Together with the DEB temperatures, one can infer their luminosity and hence obtain the distance to the system, without assuming any particular value of $G$. The estimated distance to the LMC is therefore not affected by a modified gravitational constant.

\subsection{Cepheids}
Cepheid pulsation periods are proportional to their luminosities, making them standardizable candles (after correcting also for their colour and potentially metallicity). The pulsation period of Cepheids is dictated by processes in the envelope of the star. If the envelope is unscreened (i.e., having $G > \GN$), the dynamics driving the pulsation is modified as the free-fall time is reduced by a factor $\sqrt{\GN/G}$ \citep{1980tsp..book.....C}. We assume that the pulsation period is reduced by the same factor, an approximation confirmed by models using the linear adiabatic wave equation \citep{Sakstein:2013pda}. In galaxies where Cepheid envelopes are unscreened, the PLR is shifted compared with the galaxies where Cepheid envelopes are screened.
This shift has the same effect as an increase in the Cepheid luminosity,\be
\label{eq:DeltaLA}
\Delta\log_{10} L =\frac{A}{2}\log_{10} \left(1+\frac{\Delta G}{G_{\rm N}}\right).
\ee
Here, we adopt $A=1.3$ in line with ref.~\cite{Desmond_2019}. The value for $\relG$ that goes into eq.~\eqref{eq:DeltaLA} is computed as a weighted average of $\relG$ throughout the Cepheid, where the weighting function is obtained from ref.~\cite{1950ApJ...112....6E}, prioritizing regions which are more important in driving the pulsations.

The luminosity of a Cepheid is due to hydrogen burning in a thin shell outside the helium core. If the thin shell is unscreened, it must burn more fuel to balance the increased gravitational field, thus being more luminous. In this case, a modified stellar structure code \citep{Sakstein:2019qgn} can be used to derive,
\be
\label{eq:DeltaLB}
\Delta\log_{10} L = B\log_{10} \left(1+\frac{\Delta G}{G_{\rm N}}\right),
\ee
where $B$ is determined by the mass of the
Cepheid and depends on whether it is at the second or third crossing of the instability strip. Generally $B\sim 4$ which we adopt here. The value of $\relG$ that goes into eq.~\eqref{eq:DeltaLB} is the value in the thin shell surrounding the core.

To summarize, there are two fifth force effects contributing to a shift in the Cepheid PLR. The first due to modified dynamics in the envelope and the second due to a modified burning rate close to the core. Since the envelope and the core represent different environments, we expect generically different values for $\relG$ in the envelope and close to the core. For chameleon \citep{Khoury:2003aq}, symmetron \citep{Hinterbichler:2010es}, and similar models exhibiting a thin-shell screening mechanism, the Cepheid core is more screened than the envelope.

\subsection{Type Ia supernovae}
\label{sec:DeltaMSN}
If the gravitational force is effectively increased due to the presence of a fifth force in an unscreened white dwarf, its Chandrasekhar mass will be reduced. This results in less fuel available for a possible SNIa explosion. Together with the modified Ni$^{56}$ mass and the standardization of the luminosity, it results in a change of the SNIa absolute magnitude,
\be
\label{eq:MSNmod}
\Delta M_{\rm SN}=-2.5 C\log_{10} \left(1+\frac{\Delta G}{G_{\rm N}}\right),
\ee
with $C=1.46$ \citep{Wright:2017rsu}. The inferred value of $H_0$ will decrease if Hubble flow SNIa are more unscreened than those in the host galaxies.

Since white dwarfs are much denser than Cepheids, the relative increase in the gravitational constant $\relG$ in eq.~\eqref{eq:MSNmod} is negligible in the range where the fifth force affects the Cepheids in the calibration of $H_0$. For example, for a symmetron model with a Compton wavelength of $\lC = 10 \, \mathrm{Mpc}$ and coupling constant $g=1$, typically $\relG \lesssim 10^{-6}$.

\subsection{Consistency test}
\noindent \textbf{TRGB.} 
When the hydrogen at the core of a solar mass star is exhausted, energy will mainly be generated by hydrogen fusion in a shell around the core.
As the pressure and temperature of the core increases, for stars with masses less than $1.8\, M_\odot$, helium will undergo a rapid nuclear fusion process -- the helium flash -- resulting in a break in the luminosity evolution of the star. This discontinuity is called the tip of the red-giant branch (TRGB). When measured in the near infrared $I$-band ($\sim 800$ nm), this tip is a standard candle with an absolute magnitude of $M_I\simeq -4.0$. The TRGB can be used as an alternative way of constructing a distance ladder, ultimately leading to a value of $H_0$ as in ref.~\cite{Desmond:2020wep}. Here however, we use it as a consistency test, comparing Cepheid and TRGB distance estimates to the same galaxies, as in ref.~\cite{Desmond_2019}. 
We make use of TRGB data from ref.~\cite{Freedman_2019}, where the TRGB distance from the galaxy centers are estimated from the images of the TRGB galaxy fields and the velocity dispersion are obtained from the HI linewidth W20 as tabulated in the Extragalactic Distance Database (EDD). Given that the TRGB distance estimates in ref.~\cite{Freedman_2019} are calibrated using the TRGB in the LMC, we include screening effects in the LMC, estimating TRGB galactocentric distances using the TRGB galaxy fields depicted in 
ref.~\cite{2012AcA....62..247U} along with the exclusion region described in ref.~\cite{Freedman_2019}.

The luminosity of the RGBs is determined by a thin shell of hydrogen surrounding the helium core. A modified gravitational constant in this region results in a distance modification which is well-fit by \citep{PhysRevD.101.129901},
\be
\label{eq:TRGBdistMod}
\frac{D_{\rm true}}{D_{\rm GR}}=1.021\sqrt{1-0.04663\left(1+\frac{\Delta G}{\GN}\right)^{8.389}}.
\ee
To estimate $\relG$, we assume fiducial values of the mass and radius of the RGBs: $M = 0.6M_\odot$ and $R = 207R_\odot$. When $\lC = 1 \, \mathrm{Mpc}$, typically $\relG \ll 10^{-6}$, but with $\relG$ varying over many orders of magnitude depending on the velocity dispersion of the host galaxy and the distance of the RGBs from the center of the galaxy. For $\lC = 10 \, \mathrm{Mpc}$, $\relG \sim 0.1$, varying one order of magnitude up or down between RGBs in different galaxies. When $\lC  = 100 \, \mathrm{Mpc}$, $\relG \simeq 1.9$. Here, we have assumed that $g = \sa = 1$.

Since the Cepheid and TRGB distance modifications have opposite directions, they should disagree systematically in the presence of significant fifth forces, making it possible to constrain the size of the effect.

\section{Results}
\begin{figure*}[t]
    \centering
    \includegraphics[width=0.49\linewidth]{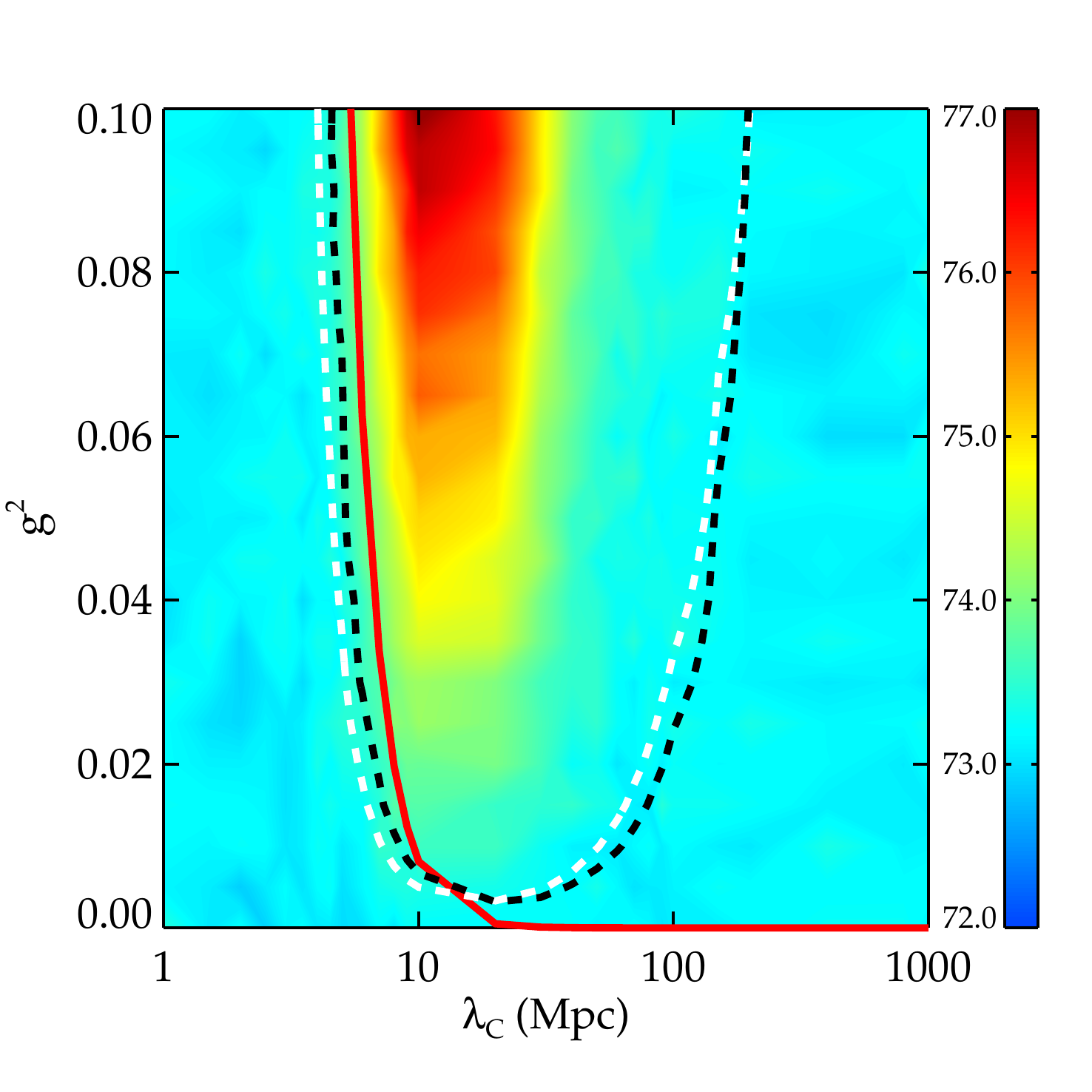}
    \includegraphics[width=0.49\linewidth]{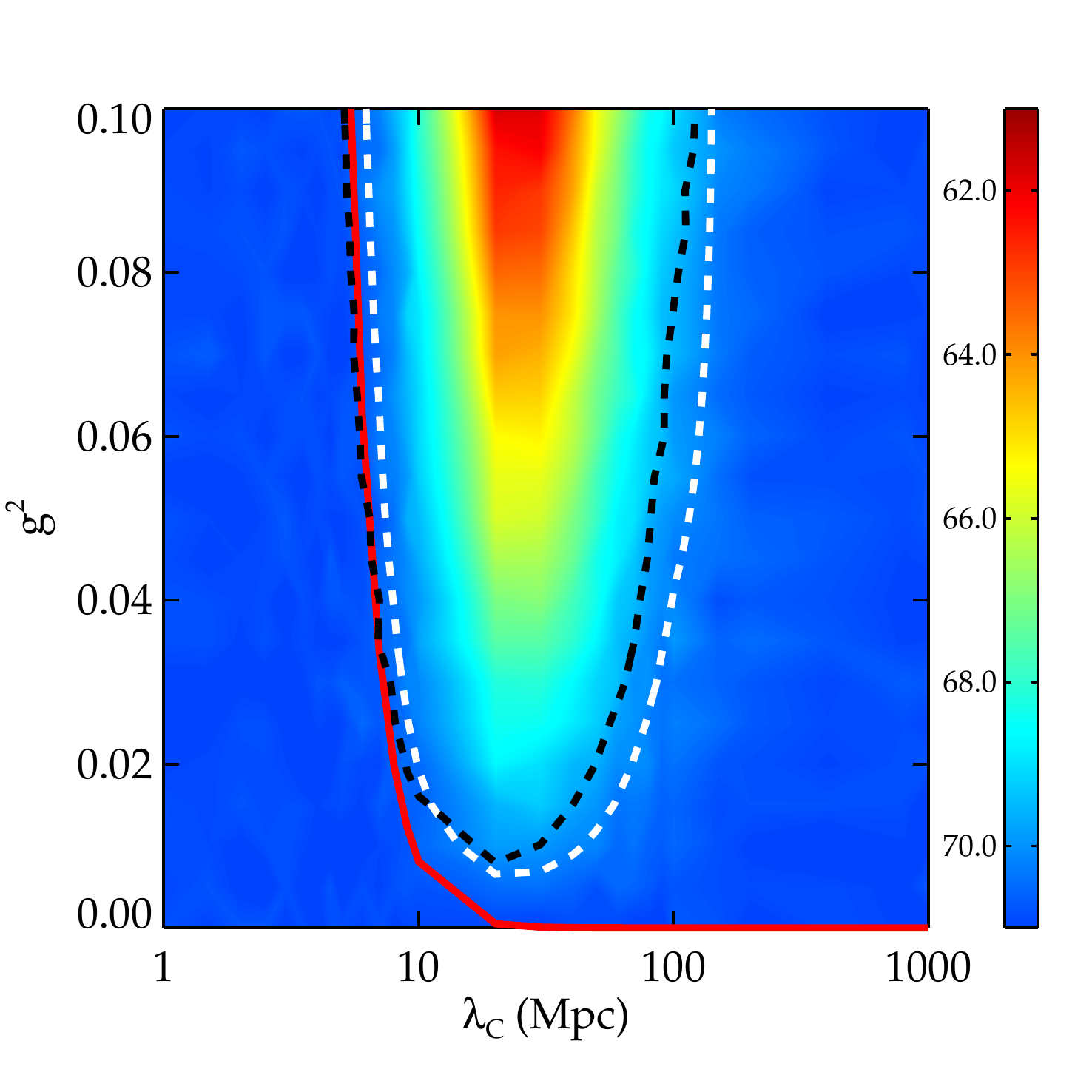}
    \caption{Derived $H_0$ as a function of $\lC$ and $g$ for the symmetron model. The color coding indicates the value of $H_0$ (note the difference in scales between the left and right panels). The dashed curves indicate $68.3\%$ confidence contours such that everything above them are excluded by the cosmic distance ladder data to this level. Black curve: constraints from the local Cepheid/SNIa distance ladder. White curve: tension between Cepheid-based and TRGB distances. Red curve: solar system constraints. \emph{Left}: including all anchor galaxies (MW, LMC, and N4258). For all theory parameters, $H_0$ is greater than or equal to the SH0ES value $H_0 = 73.0 \, \mathrm{km/s/Mpc}$, worsening the tension with Planck data. \emph{Right}: using only N4258 as anchor galaxy. In this case, the value of $H_0$ decreases, easing the Hubble tension. Values down to $H_0 = 69.6 \, \mathrm{km/s/Mpc}$ are allowed.}
    \label{fig:symk2vslambdaAll}
\end{figure*}

\begin{figure*}[t]
    \centering
    \includegraphics[width=0.45\linewidth]{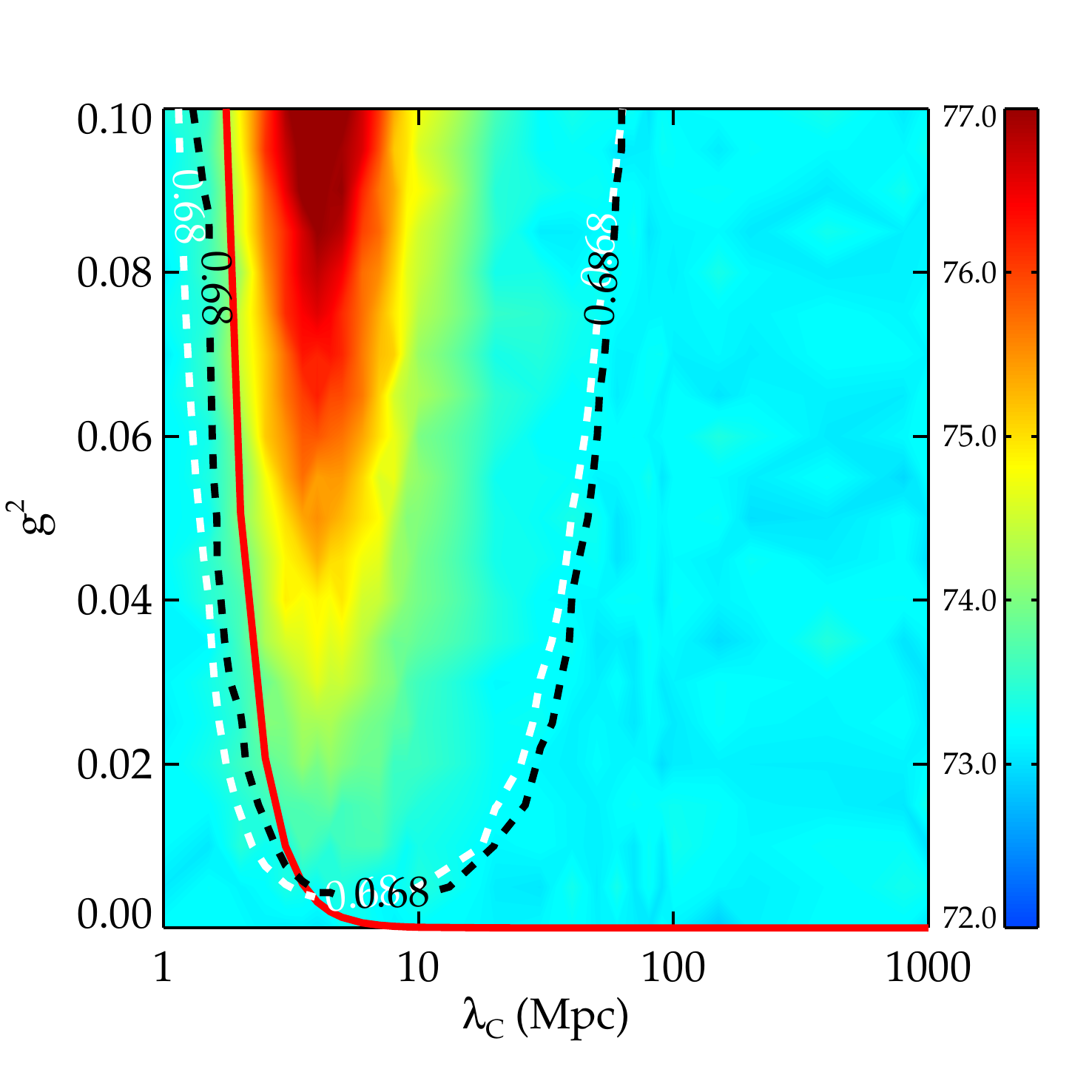}
    \includegraphics[width=0.45\linewidth]{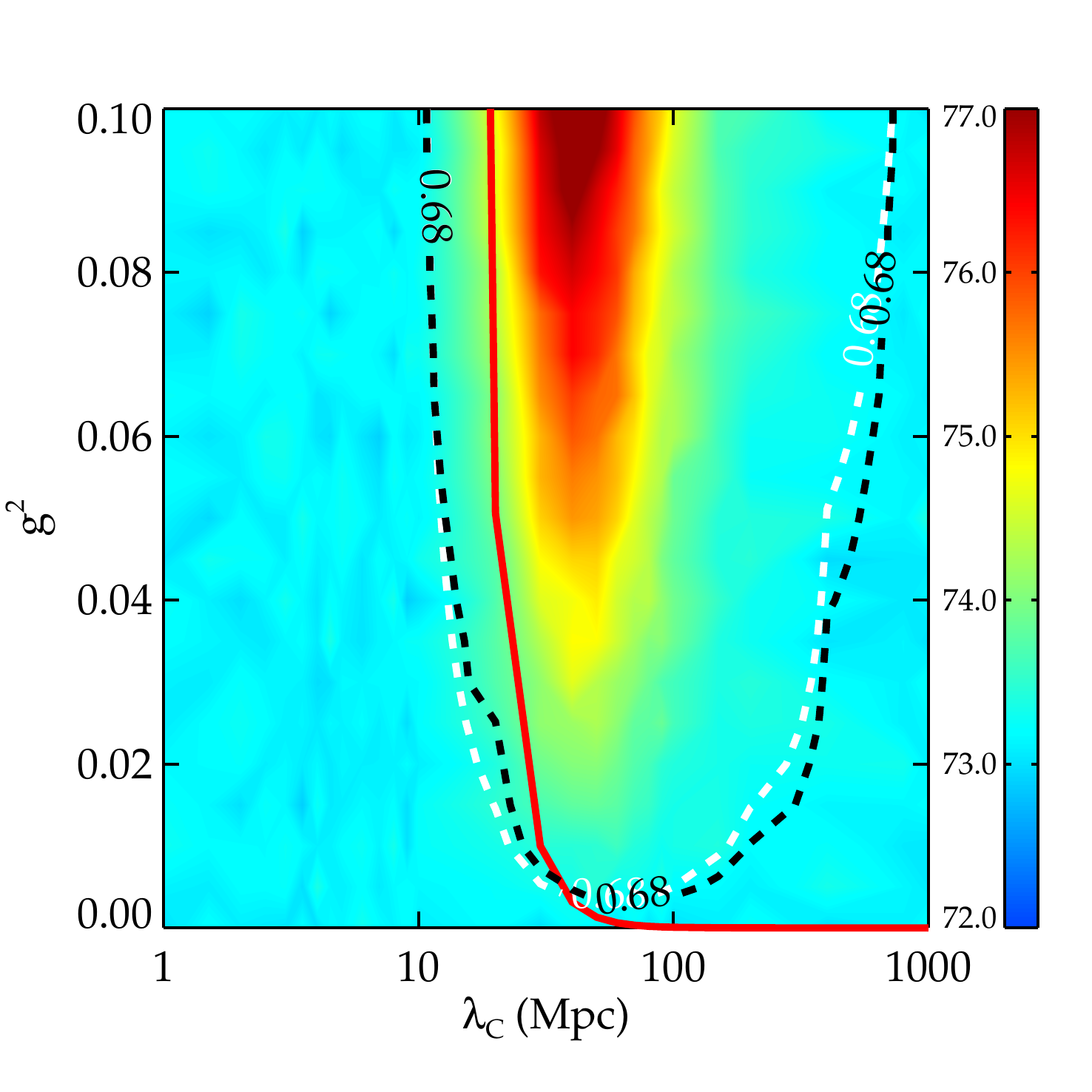}
    \caption{Derived $H_0$ as a function of $\lC$ and $g$ for the symmetron model. All anchor galaxies included. The color coding indicates the value of $H_0$. The dashed curves indicate $68.3\%$ confidence contours such that everything above them are excluded by the cosmic distance ladder data to this level. Black curve: constraints from the local Cepheid/SNIa distance ladder. White curve: tension between Cepheid-based and TRGB distances. Red curve: solar system constraints. \emph{Left}: $\sa = 0.1$. \emph{Right}: $\sa = 10$.}
\label{fig:symk2vslambda_noMCMC_sa01All}
\end{figure*}

\begin{figure*}[t]
    \centering
    \includegraphics[width=0.45\linewidth]{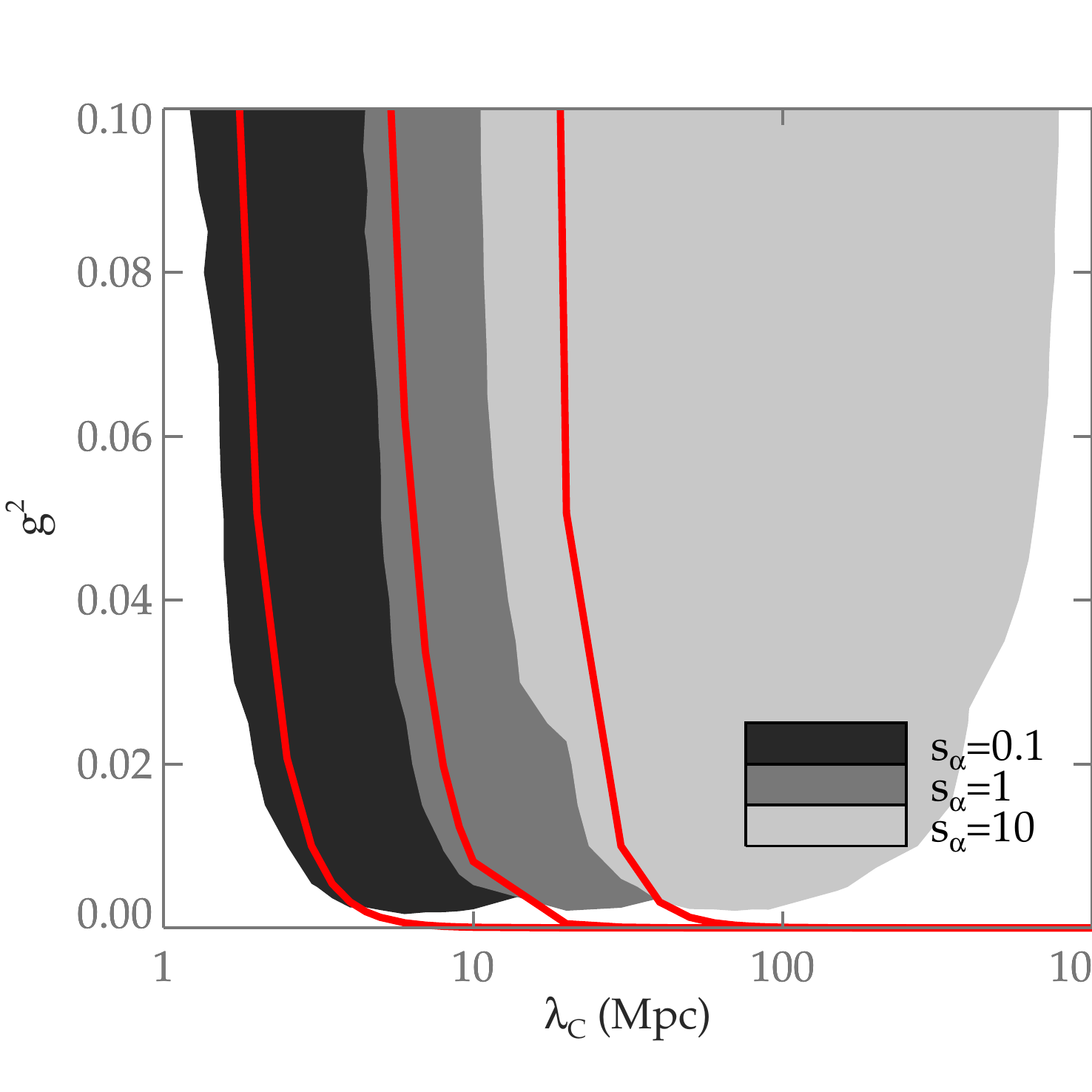}
    \caption{Exclusion plot in the symmetron parameter space for three different values of $\sa$. The gray scale regions are excluded based on cosmic distance ladder data. The red curves represent the corresponding contours for solar-system test such that, for each respective value of $\sa$, everything above them is excluded. We see that cosmic distance ladder constraints are competitive with solar-system tests at the lower end of $\lC$.}
\label{fig:symk2vslambda_combinedAll}
\end{figure*}

Fig.~\ref{fig:symk2vslambdaAll} summarizes the main result. The color coding indicates the inferred value of $H_0$ by the Cepheid/SNIa-based distance ladder, assuming a symmetron model with $\sa = 1$. As can be seen, the inferred value of $H_0$ with the symmetron model is always greater than the SH0ES value unless $\lC$ is very large or very small or if $g$ is small, in which case we approach the SH0ES value $H_0 = 73.0 \, \mathrm{km/s/Mpc}$. Hence, the symmetron model does not solve the Hubble tension, rather the opposite. As we argue in the Discussion, this is explained by the fact that the anchor galaxies span almost the same range of velocity dispersions as the host galaxies, which means that the anchor galaxies are not unusually screened compared with the host galaxies. The only possibility of easing the tension is to use N4258 as the only anchor galaxy, since it is unusually screened compared with most host galaxies. The result of this scenario is presented in the right panel of Fig.~\ref{fig:symk2vslambdaAll}. In this case, values down to $70 \, \mathrm{km/s/Mpc}$ are allowed, easing the tension to some degree. In Appendix~\ref{sec:IndAnch}, we show the results when using the MW or the LMC as the only anchor galaxy. 

Interestingly, for the model to be consistent with the distance ladder data, we obtain significant constraints on the theory. These constraints are independent of solar-system tests. The dashed curves in Fig.~\ref{fig:symk2vslambdaAll} indicate $68.3\,\%$ confidence contours such that everything above it is excluded to this level. 
The black curve corresponds to the local Cepheid/SNIa distance ladder, including the uncertainty in the physical position of the Cepheids as well as the uncertainty in the velocity dispersion of the galaxies. The white curve is derived comparing Cepheid and TRGB distances. Notably, it follows the distance ladder constraints rather closely.

In Fig.\ref{fig:symk2vslambdaAll}, we also include the parameter space excluded by solar-system tests which are currently the most constraining tests of the symmetron model.\footnote{Imposing solar-system tests, the requirement eq.~\eqref{eq:SolSystConstr} limits the energy scale of the symmetron potential to such an extent that it cannot act as dark energy \citep{Hinterbichler:2011ca}. Accordingly, under these constraints, the symmetron model does not influence the value of the Hubble constant inferred from CMB data.} In ref.~\cite{Hinterbichler:2010es}, the authors analyze constraints from solar-system tests and binary pulsars. Assuming $g \simeq 1$, they infer $M \lesssim 10^{-3}\mpl$ or, equivalently,
\begin{equation}
\label{eq:SolSystConstr}
    \lC \lesssim 3 \sqrt{\sa} \, \mathrm{Mpc}.
\end{equation}
To facilitate comparison with the full constraints from the cosmic distance ladder in Fig.~\ref{fig:symk2vslambdaAll}, we generalize the analysis in ref.~\cite{Hinterbichler:2010es} by allowing for a general value of $g$, see Appendix~\ref{sec:SolarSystTests} for details. The result is,
\begin{equation}
    g \lesssim \left( \frac{3 \sqrt{\sa} \, \mathrm{Mpc}}{\lC}\right)^2.
\end{equation}
From Fig.~\ref{fig:symk2vslambdaAll}, we conclude that the observational constraints from the cosmic distance ladder is not only a new and independent probe of the theory but also competitive with respect to the its constraining power in the parameter space.

In particular, in the standard coupling case when $g = \sa = 1$, the region $\lC \gtrsim 3 \, \mathrm{Mpc}$ is excluded by solar-system tests. This should be compared with ${2.5 \, \mathrm{Mpc} \lesssim \lC \lesssim 400 \, \mathrm{Mpc}}$ being excluded by the cosmic distance ladder at $95\%$ confidence level in the standard coupling case.\footnote{In our analysis, the exact numbers for the $95\%$ confidence level are limited by the grid density in the parameter space. Close to the lower limit ($2.5 \, \mathrm{Mpc}$), the grid spacing is $0.5 \, \mathrm{Mpc}$ and close to the upper limit ($400 \, \mathrm{Mpc}$), the grid spacing is $100 \, \mathrm{Mpc}$.} The upper bound on $\lC$ from solar-system tests coincides with the region of the parameter space where we have the greatest effect of the fifth force on the distance ladder and thus the strongest constraints from these data. This is due to the fact that the solar-system constraints coincide with the requirement that the MW is on the edge of being screened.

In Fig.~\ref{fig:symk2vslambdaAll}, we have assumed $\sa = 1$. 
In Fig.~\ref{fig:symk2vslambda_noMCMC_sa01All}, we show the same results for $\sa = 0.1$ and $\sa = 10$. In Fig.~\ref{fig:symk2vslambda_combinedAll}, we show a combined exclusion plot, including the cases $\sa = 0.1$, $\sa = 1$, and $\sa=10$ together with the solar-system tests for the same parameters. Decreasing $\sa$ has the same effect as an overall decrease in the energy density by the same factor [cf. eq.~\eqref{eq:alpha}]. In this case, all galaxies are more unscreened, effectively shifting the ``U-shaped'' confidence contour towards the left. Correspondingly, increasing $\sa$ shifts the confidence contour to the right. The amount of shift of the contours towards the left/right scales roughly as $\sqrt{\sa}$ (cf. the discussion about the scaling of the thin-shell mechanism in Section~\ref{sec:Discussion}).

\section{Discussion}
\label{sec:Discussion}
In ref.~\cite{Hinterbichler:2011ca}, they show that $\alpha$ is inversely related to the so-called thin-shell factor,
\begin{equation}
\label{eq:thinshellfactor}
    \relR = \frac{1}{\alpha},
\end{equation}
where $\alpha$ is the dimensionless density, defined in eq.~\eqref{eq:alpha}. As the name suggests, when $\Delta R/R \ll 1$, there is a thin shell at the radius of the source where the scalar field transits from zero to the VEV, similar to the chameleon mechanism. For the symmetron model, the thin-shell factor depends both on the theory parameters $\sa$ and $\lC$, as well as the density of the source, cf. \eqref{eq:alpha}. When $\Delta R/R \ll 1$, the strength of the fifth force relative to the Newtonian force is,
\begin{equation}
\label{eq:relGthinshell}
    \frac{F_5}{\FN} = \frac{\Delta G}{\GN} \simeq 6g^2 \psi_\mathrm{env}^2 \relR,
\end{equation}
and thus the fifth force is screened (unless $g \gg 1$), see for example ref.~\cite{Hinterbichler:2011ca}. Here, $\psi_\mathrm{env}$ is the value that the scalar field approaches faraway from the source. In Appendix~\ref{sec:solveEoM}, we show in detail how we solve the EoM.

Since we can write eq.~\eqref{eq:thinshellfactor} as,
\begin{equation}
\label{eq:thinshellfactor2}
    \relR = \frac{1}{3 \sa} \lChat^2 \frac{\rho_c}{\rho},
\end{equation}
the thin-shell mechanism is effective when,
\begin{equation}
    \lChat^2 \frac{\rho_c}{\rho} \ll 3 \sa.
\end{equation}
In this case, the source is screened. For a galaxy of (halo) radius $R \sim 100 \, \mathrm{kpc}$, this happens when $\lC \ll 3 \sqrt{\sa} \, \mathrm{Mpc}$.

On the other hand, since $\Delta R/R \gg 1$ when $\lC \gg 3 \sqrt{\sa} \, \mathrm{Mpc}$, there is a gradual increase of the value of the scalar field from the region inside the source to the region outside the source and the fifth force strength approaches values of,
\begin{equation}
\label{eq:relGsmallAlpha}
    \frac{F_5}{\FN} = \frac{\Delta G}{\GN} \simeq 2 g^2 \psi_\mathrm{env}^2, \quad \relR \gg 1,
\end{equation}
see for example ref.~\cite{Hinterbichler:2011ca}. In this case, the fifth force is unscreened and there is no thin-shell mechanism at work.

From eq.~\eqref{eq:rhoIstohmlSph} and eq.~\eqref{eq:thinshellfactor2}, we see that the thin shell factor is inversely proportional to $\rho R^2 \propto \sigma^2$ for a source with an isothermal sphere profile (assuming $R_{300} \gg r_0$). This means that galaxies with small velocity dispersion have a less pronounced thin-shell mechanism and thus are more unscreened compared with galaxies with large velocity dispersion. This implies that the relative spread of $\relG$ is greater in galaxies with large $\sigma$ where $\relG$ has a strong distance-dependence. In absolute terms however, this effect is counteracted by the fact that $\relG$ is suppressed in galaxies with large $\sigma$.

\begin{figure*}[t]
    \centering
    \includegraphics[width=\linewidth]{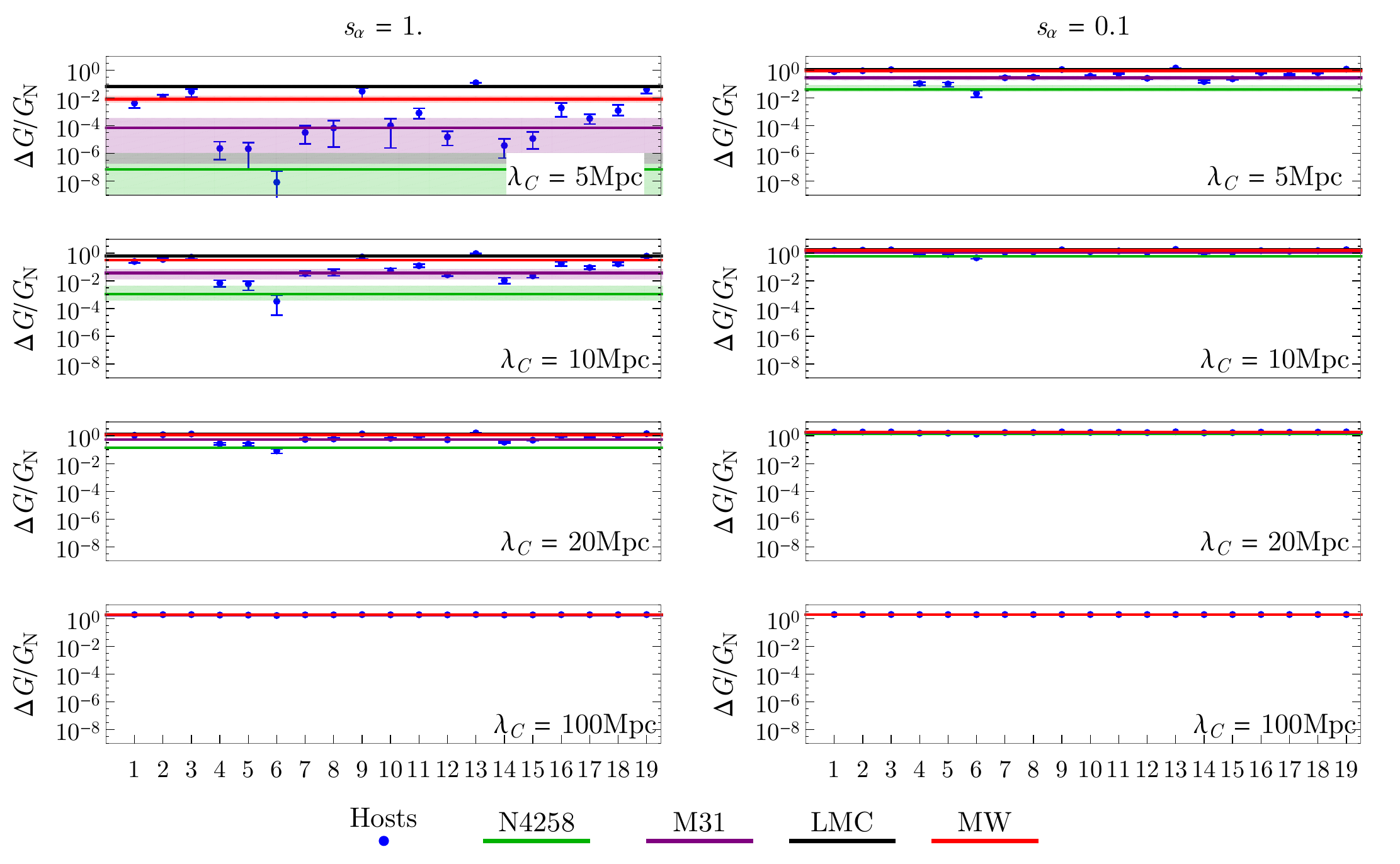}
    \caption{Mean value of $\relG$ for each galaxy. The mean is calculated with respect to all distance ladder Cepheids in the galaxy. Here, we calculate $\relG$ using $r = \pi R/2$ and $\sigma = \sigma_\mathrm{mean}$, see Section~\ref{sec:CalcRelG}. The ``error bars'' denote the largest and smallest value assumed by the Cepheids in that galaxy. The numbering of the host galaxies is the same as in Tab.~\ref{tab:sigmalist}. From top to bottom, the Compton wavelengths are: $\lC=5 \, \mathrm{Mpc}$, $\lC=10 \, \mathrm{Mpc}$, $\lC=20 \, \mathrm{Mpc}$, $\lC=100 \, \mathrm{Mpc}$. \emph{Left}: $\sa = 1$. \emph{Right}: $\sa = 0.1$. When the Compton wavelength is small, $\relG$ spans several orders of magnitude between the different galaxies. Note however that for all of them, $\relG$ is small. When the Compton wavelength is large, $\relG \simeq 2g^2$ for all galaxies.}
    \label{fig:lCrange}
\end{figure*}

\begin{table*}[t]
  \centering
  \caption{The host galaxies are listed and numbered in the same order as in Fig.~\ref{fig:lCrange}. The anchor galaxies are listed on the last row. Velocity dispersions are obtained from the EDD.}
  \label{tab:sigmalist}
  \begin{tabular}{r|cccccccccc}
    \hline\hline 
    Number: &  1 & 2 & 3 & 4 & 5 & 6 & 7 & 8 & 9 & 10\\
    Name: &  M101 & N1015 & N1309 & N1365 & N1448 & N2442 & N3021 & N3370 & N3447 & N3972\\
     $\sigma$/km/s:  & $194 \pm 5$ & $188 \pm 10$ & $161 \pm 6$ & $404 \pm 5$ & $414 \pm 7$ & $514 \pm 21$ & $291 \pm 8$ & $287 \pm 12$ & $152 \pm 6$ & $266 \pm 11$\\
     \hline \hline
     Number: & 11 & 12 & 13 & 14 & 15 & 16 & 17 & 18 & 19\\
     Name: & N3982 & N4038 & N4424 & N4536 & N4639 & N5584 & N5917 & N7250 & U9391\\
     $\sigma$/km/s:  & $232 \pm 7$ & $294 \pm 8$ & $95 \pm 5$ & $353 \pm 6$ & $303 \pm 7$ & $215 \pm 5$ & $237 \pm 6$ & $203 \pm 10$ & $140 \pm 15$\\
     \hline\hline
     Name: & MW & LMC & N4258 & M31\\
     $\sigma$/km/s: & $169 \pm 50$ & $10 \pm 2$ & $442 \pm 5$ & $301 \pm 1$\\
\hline\hline
  \end{tabular}
\end{table*}

These results are exemplified in Fig.~\ref{fig:lCrange}. As expected, $\relG$ is greater in less massive galaxies (i.e., galaxies with small velocity dispersion). However, for large values of the Compton wavelength (such as $\lC = 100 \, \mathrm{Mpc}$ in Fig.~\ref{fig:lCrange}), the thin-shell mechanism is broken in all galaxies and the fifth force approaches the same value, $\relG \simeq 2g^2 = 2$, in all galaxies. Accordingly, the spread in $\relG$ is small, as indicated by the ``error bars'' in the figure.
For these Compton wavelengths, the distance dependence of $\relG$ is the strongest in the galaxies with the largest values of $\sigma$ (such as N4258). 

For smaller values of the Compton wavelength (cf. $\lC = 5 \, \mathrm{Mpc}$ in Fig.~\ref{fig:lCrange}), $\relG$ is smaller and the thin-shell mechanism becomes effective. The largest galaxies (such as the N4258) exhibit pronounced thin-shell mechanisms and thus have the strongest distance dependence of $\relG$. On the other hand, they have the smallest absolute values of $\relG$. It turns out that for these Compton wavelengths, the smaller galaxies (such as the LMC) have the largest spread in $\relG$. 

From these observations we can understand qualitatively how the fifth force affects the calibration of the cosmic distance ladder for different values of the theory parameters $(\sa,\lC,g)$. When $\lC \gg 3 \sqrt{\sa} \, \mathrm{Mpc}$, the thin-shell mechanism is broken in all galaxies and all galaxies exhibit similar values of $\relG$. Since the fifth force effect is almost the same in all galaxies, the calibration of $H_0$ is not affected in this case. When $\lC \ll 3 \sqrt{\sa} \, \mathrm{Mpc}$, all galaxies exhibit a thin-shell mechanism and, although the relative spread in $\relG$ will be large, we retain GR results with no effect on $H_0$ since the fifth force is suppressed in all galaxies.

Hence, we expect to find the greatest effect on the inferred value of $H_0$ in the intermediate region where the Compton wavelength is $\lC \sim  3 \sqrt{\sa} \, \mathrm{Mpc}$ in which case small galaxies such as the LMC have a relatively large values of $\relG$ while larger ones such as N4258 do not. These considerations explain the ``U-shaped'' confidence contours in Figs.~\ref{fig:symk2vslambdaAll}-\ref{fig:symk2vslambda_combinedAll}. 
High values of $g$ are allowed if $\lC$ is small or large enough. On the other hand, since $g$ sets the overall scale of the fifth force, all values of $\lC$ are allowed for small enough $g$.

For fixed theory parameters, the value of $\relG$ in a galaxy is mainly determined by the velocity dispersion of the galaxy. The anchor galaxies (MW, LMC, and N4258) span a large range of velocity dispersions, from $10 \, \mathrm{km/s}$ for the LMC to $442 \, \mathrm{km/s}$ for N4258. To alleviate the tension, anchor galaxies should be screened to a larger extent compared to host galaxies. If we calibrate the distance ladder using only one of the anchors, in case of the LMC, since it is unusually unscreened we get a greater value for $H_0$ compared with the nominal $73 \, \mathrm{km/s/Mpc}$ (right panel of Fig.~\ref{fig:symk2vslambdaMW}). Calibrating with only the MW, we also expect $H_0$ to be somewhat higher since it is more unscreened than most of the hosts. On the contrary, with only N4258, $H_0$ should decrease due to this galaxy being unusually screened (right panel of Fig.~\ref{fig:symk2vslambdaAll}). When using all three anchors, as evident from the left panel of Fig.~\ref{fig:symk2vslambdaAll}, the net effect is that the inferred $H_0$ increases when including screening effects

In ref.~\cite{Desmond_2019}, the authors analyze the effects of fifth forces on the calibration of the cosmic distance ladder, focusing on effective models which parameterize the strength of the fifth force using proxy fields. Their result is that the local effects of a fifth force may help solve the Hubble tension. Here instead, we focus on symmetron screening, calculating the fifth force from the theory. In this case, we show that the tension is not solved when employing all three anchor galaxies, rather it gets worse. On the other hand, fitting the symmetron model to the cosmic distance ladder data allows to set new limits on the parameter space of the theory which turn out to be competitive with solar-system tests. Obviously, this does not rule out the possibility that another screening mechanism (such as those in ref.~\citep{Desmond_2019}) could solve the tension. More specifically, to solve the tension, the anchor galaxies should (on average) be significantly more screened than the host galaxies. 

This raises the question whether we could use the cosmic distance ladder data to impose significant constraints on other screening mechanisms as well. The answer should depend on the details of the screening mechanism in question. For example, a chameleon model is already heavily constrained and we do not expect the cosmic distance ladder to be competitive \citep{Burrage:2016bwy,Desmond_2019}. In a follow-up paper, we plan to evaluate this in detail for a larger number of models. 

We have analyzed how a fifth force affects the Cepheids and TRGBs when calibrating the cosmic distance ladder. Since WDs are more compact than Cepheids, we expect the effect to be less important in the range where the fifth force has a significant impact on the Cepheids and TRGBs. For a WD far out in the halo (where the fifth force is the strongest) in a galaxy with $\sigma = 200\, \mathrm{km/s}$, $\relG < 1 \, \%$ even for $\lC = 50 \, \mathrm{Mpc}$. Here, we have set $\sa = g = 1$ and set the mass to the Chandrasekhar mass $M \simeq 1.4M_\odot$ and the radius to $R = 0.01R_\odot$ \citep{1979ApJ...228..240S}. Therefore, we 
assume that the fifth force does not affect the SNIa magnitudes and set $\Delta m_\mathrm{SN} = 0$ (cf. eq.~\eqref{eq:H0effects}).

\section{Conclusions}
In this paper, we analyze the effects of a symmetron fifth force on the calibration of the cosmic distance ladder. In particular, we infer the value of the Hubble constant from a full statistical data analysis to see whether the tension between this local value and \emph{Planck} can be alleviated. To calculate the strength of the fifth force, we solve the symmetron equations of motion for each source in the cosmic distance ladder.

In addition to considering standard uncertainties (e.g. related to Cepheid and SNIa photometry), we also account for the uncertainty in the physical distance of the Cepheids from the galactic center and the uncertainty in the velocity dispersion of the galaxies. These factors affect the strength of the fifth force acting on each Cepheid.

By employing a state-of-the-art statistical analysis, we infer the value of the Hubble constant and find that the tension between the local value and \emph{Planck} is actually worsened by the inclusion of this fifth force. Specifically, for the symmetron model, we find that ${H_0 \gtrsim 73 \, \mathrm{km/s/Mpc}}$. We base our analysis on the MW, the LMC, and N4258 as anchor galaxies. Notably, with N4258 as the only anchor galaxy, the value of $H_0$ aligns more closely with the \emph{Planck} value. This is due to the fact that N4258 is an unusually screened galaxy.

It is important to emphasize that these conclusions are specific to the symmetron model, and we do not exclude the possibility that another type of fifth force might provide a resolution to the tension. Numerous other models warrant their own dedicated analyses, which should be the subject of future work.

We fit the symmetron model to the data from the cosmic distance ladder and obtain new constraints on the model. The results are summarized in Fig.~\ref{fig:symk2vslambdaAll}. In the small $\lC$ limit, the fifth force becomes suppressed and $\relG \simeq 0$. In the large $\lC$ limit, the fifth force takes the same value ($2g^2$) for all sources, hence not affecting the calibration of the distance ladder. This explains the distinctive ``U-shape'' of the excluded region in the figure,  indicating that any value of the coupling constant $g$ is permissible for sufficiently small or large Compton wavelengths.

For the standard coupling case ($g = \sa = 1$), the distance ladder data excludes a range of Compton wavelengths, with $2.5 \, \mathrm{Mpc} \lesssim \lC \lesssim 400 \, \mathrm{Mpc}$, at a confidence level of $95 \, \%$.\footnote{The grid size is $0.5 \, \mathrm{Mpc}$ at the lower limit and $100 \, \mathrm{Mpc}$ at the higher limit, so this exclusion region is approximate.} The lower limit is competitive with the results of solar-system tests, which are currently the most constraining tests of this model.

As a consistency check, we compare distances to the same galaxies using both the Cepheid-based distance ladder and a TRGB-based distance ladder. Since the modifications in distance move in opposite directions for these two methods, it imposes additional constraints on the symmetron model. Notably, these constraints closely resemble those obtained from Cepheid-only data, as illustrated in Fig.~\ref{fig:symk2vslambdaAll}.

In conclusion, we show that the symmetron model does not resolve the Hubble tension. Nevertheless, the cosmic distance ladder data proves to be a novel and powerful tool for testing this model as well as similar theories.

\begin{acknowledgements}
EM acknowledges support from the Swedish Research Council under Dnr VR 2020-03384. Thanks to Harry Desmond for sharing data files from ref.~\cite{Desmond_2019}. Thanks to an anonymous referee for insightful comments.
\end{acknowledgements}

\appendix
\section{Solving the equations of motion}
\label{sec:solveEoM}

\subsection{Galaxies}
\label{sec:AppGalaxies}
As an approximation for the mass density of a galaxy, we assume an isothermal sphere with a core,
\begin{equation}
\label{eq:rhoIstohmlSph}
    \rho(r) = \frac{\sigma^2}{2\pi \GN} \frac{1}{r^2 + r_0^2},
\end{equation}
where $\sigma$ is the velocity dispersion and $r_0$ is the radius of the core. In this case, the EoM \eqref{eq:psiEoM} can be solved analytically if the Compton wavelength is small enough (satisfying the requirement $\lChat < 15 \sqrt{2\sa}$) and setting $r_0 = 0$. Recall that $\lChat = \lC / R_{300}$. The solution is,
\begin{subequations}
    \begin{alignat}{2}
        \psi(\rhat) &= c_1 j_n(\rhat / \lChat),& \quad \rhat &< 1,\\
        \psi(\rhat) &= c_2 \frac{e^{-\sqrt{2} \rhat/\lChat}}{\rhat} + \psi_\mathrm{env,gal},& \quad \rhat &> 1,
    \end{alignat}
\end{subequations}
where $j_n$ is the spherical Bessel function of order $n$ and,
\begin{equation}
    n = \frac{1}{2} \left[ -1 + \sqrt{1 + \left( \frac{60 \sqrt{\sa}}{\lChat}\right)^2} \right]. 
\end{equation}
The coefficients $c_1$ and $c_2$ are set by the boundary conditions and can be provided on demand. When the analytical solution is not applicable, the EoM is solved numerically using a shooting method with boundary conditions $\partial_{\rhat} \psi|_{\rhat = 0.1 \rhat_0} = 0$ where $\rhat_0$ is the radius of the core (in units of $R_{300}$) and $\psi|_{\rhat = 10} = \psi_\mathrm{env,gal}$. Here, $\psi_\mathrm{env,gal}$ is the environmental value that the scalar field of the galaxy should approach far away from the galaxy itself. For the MW, $\psi_\mathrm{env,gal}$ is set by the value of the scalar field provided by M31 and for the LMC, the environmental value is set by the value of the scalar field for the MW at the position of the LMC. For all other galaxies, we set $\psi_\mathrm{env,gal} = 1$. For the numerical solution to be well-behaved, we set a finite core radius, here $1 \, \%$ of the radius of the galaxy.

\subsection{Stars}
\label{sec:AppStars}
For simplicity, stars are modelled with a homogeneous mass density profile $\rho(r) = \mathrm{const}$. In this case, eq.~\eqref{eq:psiEoM} can be solved analytically with the result \citep{Hinterbichler:2010es},
\begin{subequations}
\label{eq:psiSol}
    \begin{alignat}{2}
    \label{eq:psiSolIn}
        \psi_\mathrm{in}(\rhat) &= \frac{c_1}{\rhat} \sinh \left( \rhat \sqrt{\alpha - \frac{1}{\lChat^2} } \right),& \quad \rhat&<1,\\
    \label{eq:psiSolOut}
        \psi_\mathrm{out}(\rhat) &= \frac{c_2}{\rhat} e^{- \sqrt{2} \, \rhat / \lChat} + \psi_\mathrm{env},& \quad \rhat&>1.
    \end{alignat}
\end{subequations}
Hats denote length scales measured in units of the radius of the source (here, the star). In \eqref{eq:psiSol}, $c_1$ and $c_2$ are integration constants set by the requirement that the field and its first derivative is continuous across the boundary $\rhat = 1$. They can be solved for analytically but the expressions are somewhat lengthy. Formally,
\begin{subequations}
    \label{eq:cSolFormal}
    \begin{align}
        \label{eq:c1SolFormal}
        c_1 &= \psi_\mathrm{env} f_1(\alpha , \lChat),\\
        \label{eq:c2SolFormal}
        c_2 &= \psi_\mathrm{env} f_2(\alpha, \lChat),
    \end{align}
\end{subequations}
for some functions $f_1$ and $f_2$. The full expressions are available on demand. Here, $\psi_\mathrm{env}$ is the environmental background value that $\psi$ approaches well outside the radius of the source (i.e., when $\rhat \gg 1$). For a star in a galaxy, this is the field value set by the galaxy itself in the neighborhood of the star. In the section above, we described how to  calculate that. Hence, the screening properties depend not only on the compactness of the source, $\alpha$, but also on the galactic environment, entering via $\psi_\mathrm{env}$. Note that for a homogeneous source (and only in this case), $\alpha$ is a constant (cf. eq.~\eqref{eq:alpha}) which can be rewritten as,
\begin{equation}
    \label{eq:alphaHom}
    \alpha = 400 \sa \, \left( \frac{\lC}{\mathrm{Mpc}}\right)^{-2} \frac{M/M_\odot}{R/R_\odot}.
\end{equation}
Inside a homogeneous, spherically symmetric source, the gravitational force (per unit mass) is,
\begin{equation}
    \label{eq:FN}
    \FN = \frac{1}{8\pi \mpl^2} \frac{M_\star(<r)}{r^2} = \frac{1}{6 \mpl^2} \rho r, \quad r < R,
\end{equation}
where $M_\star(<r)$ is the total mass enclosed within radius $r$. Using eqs.~\eqref{eq:relGdef}, \eqref{eq:alpha}, \eqref{eq:F5}, \eqref{eq:FN}, we get,
\begin{equation}
    \label{eq:DeltaG}
    \frac{\Delta G(\rhat)}{\GN} = 6g^2 \, \frac{1}{\alpha} \, \frac{1}{\rhat} \, \psi(\rhat) \, \partial_{\rhat} \psi(\rhat), \quad \rhat < 1,
\end{equation}
where $\psi(\rhat)$ is given by \eqref{eq:psiSol}. Note that $\Delta G$ depends on the radius. Outside the source,
\begin{equation}
    \FN = \frac{1}{8\pi \mpl^2} \frac{M_\star}{r^2} = \frac{1}{6 \mpl^2} \frac{\rho R}{\rhat^2}, \quad \rhat > 1,
\end{equation}
and $\relG$ becomes,
\begin{equation}
    \frac{\Delta G(\rhat)}{\GN} = 6g^2 \, \frac{1}{\alpha} \, \rhat^2 \, \psi(\rhat) \, \partial_{\rhat} \psi(\rhat), \quad \rhat > 1. 
\end{equation}
In the case of a homogeneous source with constant $\alpha$, we let $\alpha$ denote the dimensionless density inside the source even when $\rhat > 1$. Together with the solution for the scalar field eq.~\eqref{eq:psiSol} this yields the relative shift in the gravitational constant given the theory parameters $(\sa,\lC,g)$ and the radius and mass of the object (star) and the environmental value of the scalar field. The latter is determined by the galaxy hosting the star, see Section~\ref{sec:AppGalaxies}. From eq.~\eqref{eq:psiSolIn} and eq.~\eqref{eq:c1SolFormal}, we see that $\Delta G/\GN \propto \psi_\mathrm{env}^2$, hence the relative strength of the fifth force depends on the galactic value of the scalar field at the position of the star. A greater value of $\psi_\mathrm{env}$, as in vacuum where $\psi_\mathrm{env}=1$, corresponds to a stronger fifth force, as expected. In other words, for a star in a galaxy, the fifth force should be as strongest (relative to $\FN$) towards the edge of the galaxy where the average energy density is the smallest.

Setting $\psi_\mathrm{env} = 1$ and taking the limit $\lC \to \infty$ and $\alpha \to 0$ of the solution \eqref{eq:psiSol} yields an estimate of the maximum modification of $G$ that can be obtained. The result is,
\begin{equation}
\label{eq:relGmax}
    \left. \relG \right|_\mathrm{max} = 2g^2,
\end{equation}
in accordance with eq.~\eqref{eq:relGsmallAlpha}. So, a maximum effective increase in the gravitational force by a factor $1+2g^2$ can be expected in low density environments if $\lC$ is large enough.

\section{Monte Carlo sampling}
\label{sec:MonteCarlo}
The uncertainty in the physical distance of the Cepheids from the galactic center results in an uncertainty in $\relG$. The same holds for the uncertainty in the velocity dispersion of the host galaxies. In this section, we explain how we account for these uncertainties in $\relG$, using Monte Carlo methods.

\subsection{Physical distance}
\label{sec:PhysDist}
The angle $\theta_{12}$ between the galactic center and the Cepheids is calculated using,
\begin{equation}
    \label{eq:theta12}
\cos \theta_{12} =\sin \delta_1 \sin \delta_2+\cos \delta_1 \cos \delta_2 \cos(\alpha_1-\alpha_2),
\end{equation}
where $\delta = \mathrm{DEC}$ and $\alpha = \mathrm{RA}$. The 2D projected distance (which subtends the angle $\theta_{12}$ on the sky) is given by,
\begin{equation}
    R = \theta_{12} D.
\end{equation}
Here, $D$ is the angular diameter distance to the galaxy, which can be obtained from ref.~\cite{Riess:2016jrr} and is tabulated in Table~II of ref.~\cite{Desmond_2017}. Given the 2D projected distance, the physical 3D distance from the center of the galaxy to the Cepheid can be estimated using,
\begin{equation}
\label{eq:rRexpect}
    r = \frac{\pi}{2} R.
\end{equation}
This is the expectation value of $r$, assuming a spherically symmetric distribution of Cepheids in the galaxy. We stress that the exact value of $r$ is not known, except in the MW which is treated as a special case below. Eq.~\eqref{eq:rRexpect} only gives the expectation value.

To estimate the individual Cepheid distances from the MW center, we use the fact that the distance is given by the law of cosines as,
\begin{equation}
    r^2 = d_{\rm Ceph}^2+d_{\rm GC}^2-2 \, d_{\rm Ceph} \, d_{\rm GC}\cos \theta_{12},
\end{equation}
where $\cos \theta_{12}$ is given by eq.~\eqref{eq:theta12}. Here, $d_{\rm GC}=8.178$ kpc is our distance from the galactic center and $d_{\rm Ceph}$ is the distance between us and the Cepheid. The latter is estimated from the GAIA parallax according to,
\be
d_{\rm Ceph}=\frac{1}{\pi+zp},
\ee
with $\pi$ being the parallax (in mas) and $zp$ the parallax offset. Using galactic coordinates, the expression is simplified since $\delta_2=\alpha_2=0$, giving,
\be
\cos \theta_{12} = \cos \delta_1 \cos \alpha_1.
\ee
\begin{figure*}[t]
    \centering
    \includegraphics[width=0.6\linewidth]{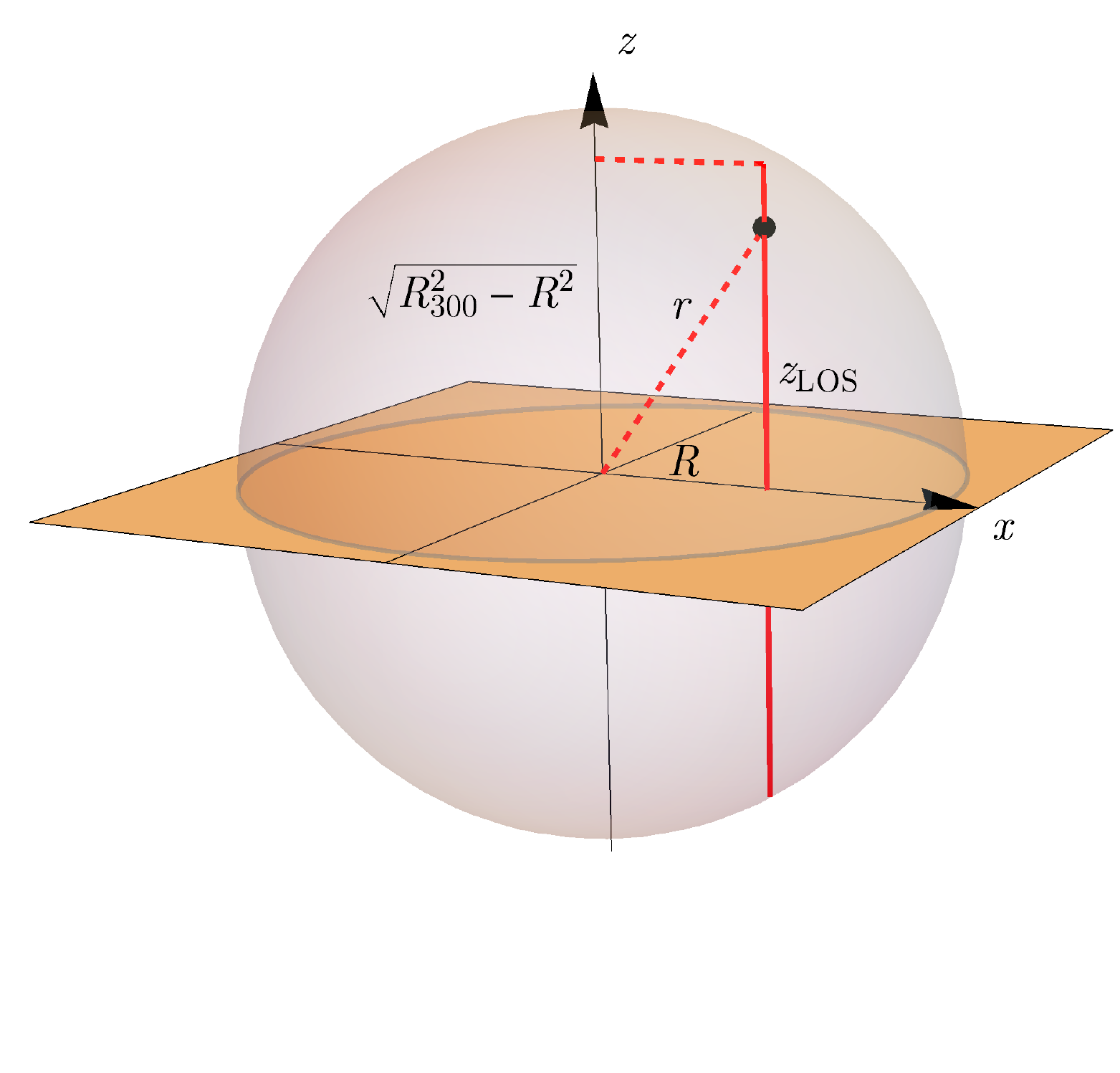}
    \vspace{-20mm}
    \caption{Geometry of a Cepheid within a galaxy of radius $R_{300}$. The projection plane is identified as the $xy$-plane (orange), $r$ is the 3D physical distance of the Cepheid (black dot) from the center of the galaxy, $R$ is the 2d projected distance, and $z_\mathrm{LOS}$ is the coordinate of the Cepheid along the line of sight, i.e. along the $z$-direction. We set $z_\mathrm{LOS}=0$ when the Cepheid lies in the $xy$-plane.}
    \label{fig:plotgeom}
\end{figure*}
When we observe a Cepheid within another galaxy, we can only infer its two-dimensional projected distance $R$ from the center of the galaxy (the projection is onto the plane normal to the line of sight (LOS)). The position of the Cepheid along the LOS, $\zlos$, is a stochastic variable that follows a probability distribution. Here, we define $\zlos=0$ as the point along the LOS where $r = R$, see Fig.~\ref{fig:plotgeom}. We assume spherical symmetry for the spatial distribution of the Cepheids and that their number density falls off as $1/r^2$. With these assumptions, the probability density function (PDF) for $\zlos$ is given by,
\begin{align}
\label{eq:PDFlos}
    &\mathrm{PDF}(\zlos) = \nonumber \\
    &\left\lbrace \begin{array}{ll}
        \cfrac{1}{2} \; \cfrac{\hat{R}}{\mathrm{arccot} \hat{R} \; (\zlos^2+\hat{R}^2)}, \quad & |\zlos| < \sqrt{1-\hat{R}^2},  \\[10pt]
        0, \quad & |\zlos| > \sqrt{1-\hat{R}^2},
    \end{array} \right\rbrace.
\end{align}
Recall, hats denote a length scale measured measured in units of $R_{300}$. The cutoff for the PDF is due to the finite extension of the galaxy, cf. Fig.~\ref{fig:plotgeom}.

\begin{figure*}[t]
    \centering
    \includegraphics[width=\linewidth]{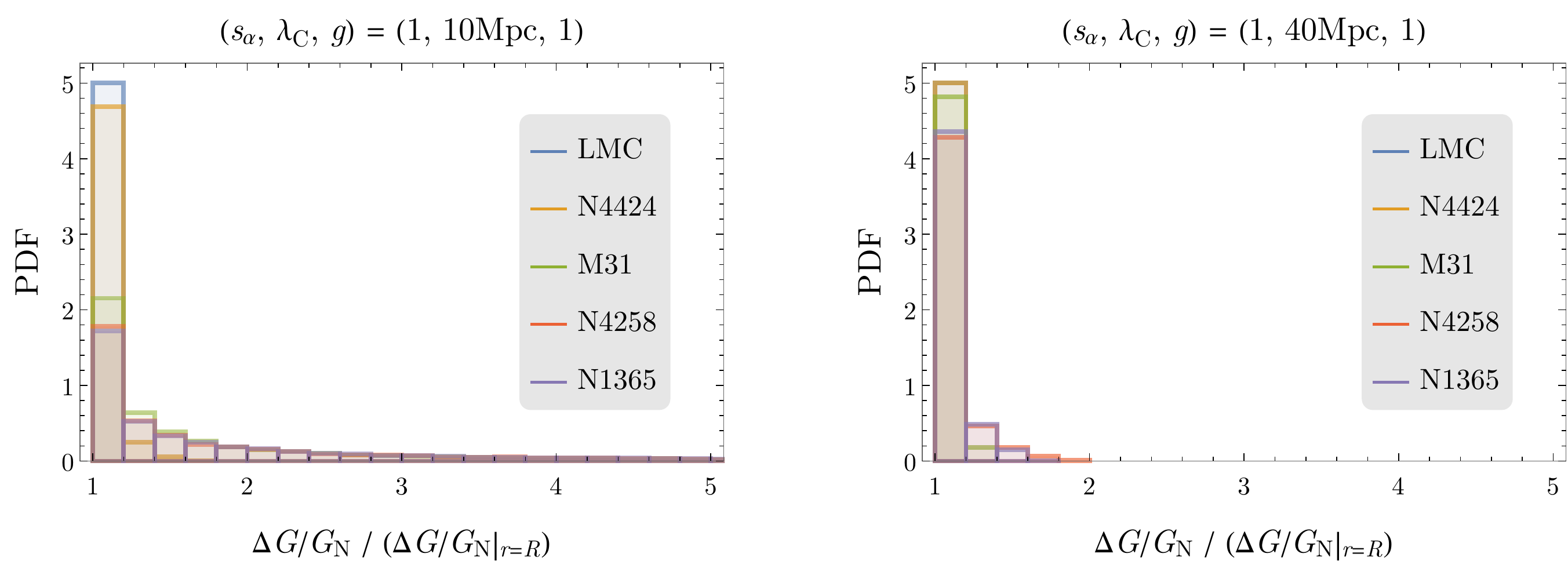}
    \caption{Example distributions for $\relG$ due to the uncertainty in the physical distance of the Cepheid from the galactic center. To generate the plots, we set the projected radius $R$ to some typical value for each galaxy and sample $10^4$ points from the distribution \eqref{eq:PDFlos}. \emph{Left}: Compton wavelength $\lC = 10 \, \mathrm{Mpc}$. In this case, more massive galaxies like N4258 and N1365 have a greater spread while small galaxies like the LMC have a very sharp peak at $\relG = \relG|_{r=R}$. \emph{Right}: $\lC = 40 \, \mathrm{Mpc}$. With a greater Compton wavelength, the spread becomes similar among all galaxies and the PDFs are narrowly concentrated around $\relG = \relG|_{r=R}$.}
    \label{fig:relGPDF_example}
\end{figure*}

\begin{figure*}[t]
    \centering
    \includegraphics[width=\linewidth]{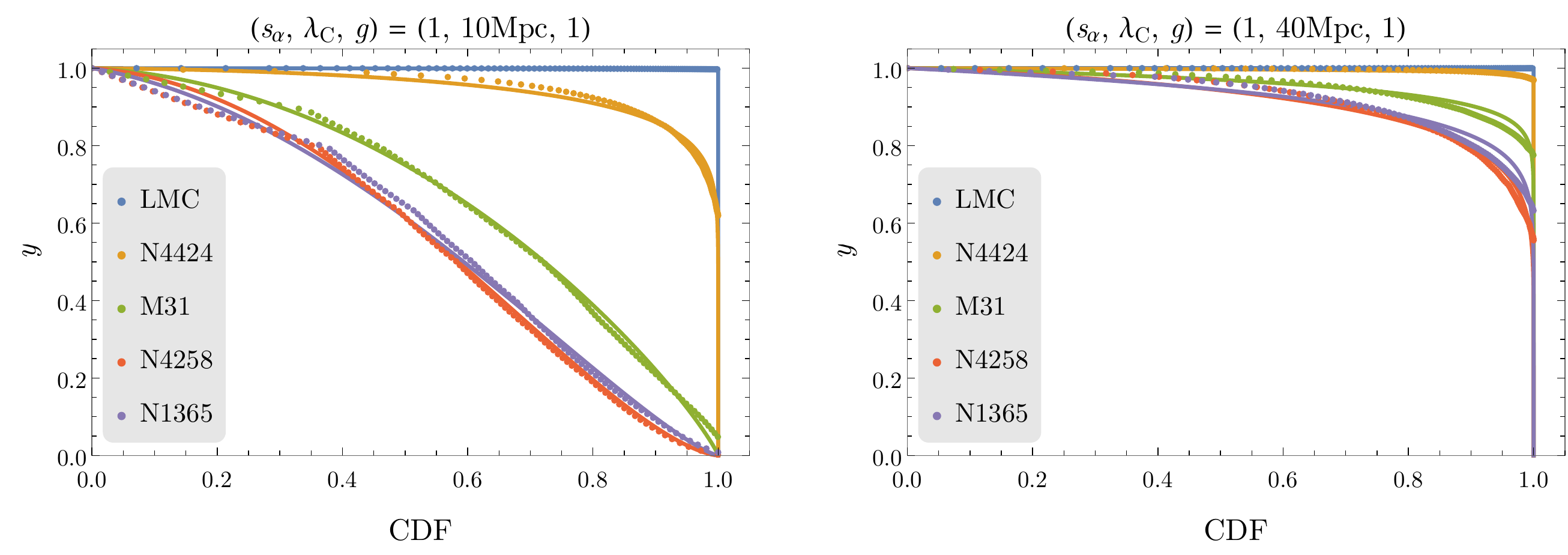}
    \caption{Example plots of $y(\mathrm{CDF})$. The radius is set to some typical Cepheid position in each galaxy. The dots mark out exact values for $y(\mathrm{CDF})$, calculated directly from the $10^4$ samples of physical distances. The solid lines represent $y(\mathrm{CDF})$ using the approximation in eqs.~\eqref{eq:yCDF}-\eqref{eq:n1n2} with the best-fit values of $(a,b,c,d)$. The exact and approximated values fit well. \emph{Left}: Compton wavelength $\lC = 10 \, \mathrm{Mpc}$. In this case, the more massive galaxies exhibit a slope in the function $y(\mathrm{CDF})$ meaning that a random realization of $y$ can take a range of values, hence the spread in $(\relG) / (\relG|_{r=R})$ is significant. \emph{Right}: Compton wavelength $\lC = 40 \, \mathrm{Mpc}$. For a larger value of $\lC$, $y(\mathrm{CDF})$ is flatter meaning that the spread in $(\relG) / (\relG|_{r=R})$ is small.}
    \label{fig:yCDF_example}
\end{figure*}

For a given Cepheid at some two-dimensional projected distance $R$, we sample $10^4$ values of $\hat{z}_\mathrm{LOS}$ from the PDF~\eqref{eq:PDFlos} and thus obtain a distribution of physical distances according to the Pythagorean theorem giving,
\begin{equation}
\label{eq:rPhysSamp}
    \frac{r}{R} = \sqrt{1 + \left(\frac{\zlos}{\hat{R}}\right)^2},
\end{equation}
cf. Fig.~\ref{fig:plotgeom}. Using the solution for $\relG$, described in Appendix~\ref{sec:AppGalaxies}-\ref{sec:AppStars}, the sample of physical distances is turned into a sample of $\relG$. For convenience, we define a ``normalized'' $\relG$ as,
\begin{equation}
\label{eq:x}
    x = \frac{\relG}{\relG|_{r=R}}.
\end{equation}
Since $\relG$ is a monotonically increasing function of $r$ and the physical distance must be greater than or equal to $R$, we must have $x \geq 1$. We now have $10^4$ samples of $x$. This can then be integrated to yield the cumulative distribution function (CDF). In Fig.~\ref{fig:relGPDF_example}, we plot the PDF for $\relG$ for some example galaxies. For convenience, we introduce the variable,
\begin{equation}
\label{eq:y}
    y = 1/x,
\end{equation}
which accordingly ranges between $0 < y \leq 1$. Inverting the CDF, we obtain the function $y(\mathrm{CDF})$. In principle, this function can be calculated for each individual Cepheid, but for efficiency we approximate it with an expression of the form,
\begin{equation}
\label{eq:yCDF}
    y(\mathrm{CDF}) = (1-\mathrm{CDF})^{n_1} (1+\mathrm{CDF})^{n_2}.
\end{equation}
For the expression to be well defined at $y=1$, we demand $n_1 >0$ and for $y$ to be a decreasing function of the CDF-value, we demand $n_1 > n_2$. Recall that we started the construction assuming a Cepheid at a specific $R$. Hence, the parameters $n_1$ and $n_2$ in the resulting $y(\mathrm{CDF})$ carry an $R$-dependence. This dependence can be approximated by,
\begin{equation}
\label{eq:n1n2}
    n_1(R) = a R^b, \quad n_2(R) = c R^d,
\end{equation}
where $(a,b,c,d)$ are constants. To calculate $(a,b,c,d)$ for a galaxy, we sample $10^4$ values of $y$ at 10 equidistant values of $R$, from the innermost Cepheid to the outermost one. From this, we obtain $y(\mathrm{CDF})$ at each of these $R$. Then we fit these exact values of $y(\mathrm{CDF})$ with the approximations in eqs.~\eqref{eq:yCDF}-\eqref{eq:n1n2} and obtain $(a,b,c,d)$ as the best-fit parameters. So, for each galaxy, the parameters $(a,b,c,d)$ define the $y(\mathrm{CDF})$ function according to eqs.~\eqref{eq:yCDF}-\eqref{eq:n1n2}. Besides being computationally efficient, this makes it simple to export the function to any desired programming language. In Fig.~\ref{fig:yCDF_example}, we show some examples of $y(\mathrm{CDF})$, both exact values and fitted. For small galaxies and large Compton wavelengths, $y$ is essentially flat, meaning that almost all random samples of $y$ equals unity. On the other hand, increasing the galaxy mass or decreasing the Compton wavelength introduces a slope in $y$ such that a random sample can take a range of values. The spread in $(\relG) / (\relG|_{r=R})$, for selected example cases, is depicted Fig.~\ref{fig:relGPDF_example}.

\subsection{Velocity dispersion}
\label{sec:VelDisp}
\begin{figure*}[t]
    \centering
    \includegraphics[width=\linewidth]{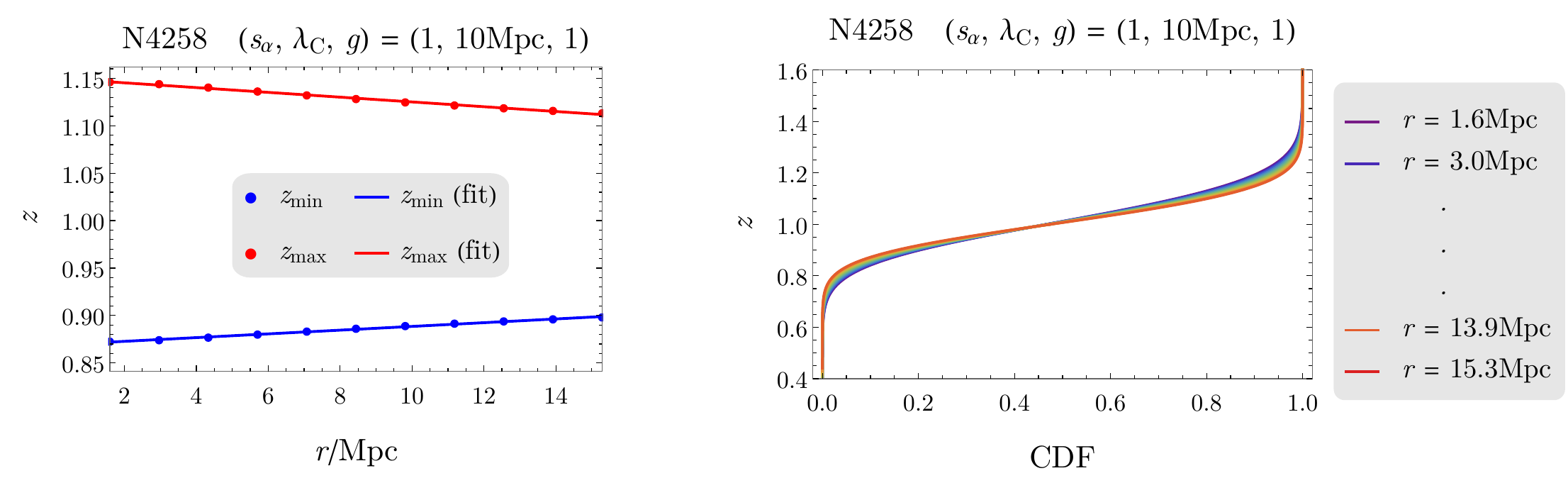}
    \caption{Example plots for $z$ for N4258 in the case $(\sa,\lC,g)=(1,10 \, \mathrm{Mpc},1)$. \emph{Left}: $\zmin$ and $\zmax$ as functions of the radius (the distance from the galactic center). Dots represent exact values and the lines are fitted. The exact values and the fits agree with excellent precision. In this example, the value of $\relG$ can deviate $\simeq 15 \%$ from the default value $\relG|_{\sigma = \sigma_\mathrm{mean}}$ within the limits $\sigma = \sigma_\mathrm{mean} \pm \Delta \sigma$. \emph{Right}: $z(\mathrm{CDF})$ for different values of the radius $r$ according to the color coding. Bluer colors denote smaller radii and vice versa for redder colors. As seen in the figure, if we sample a random number for the CDF within the range  $[0,1]$, the vast majority of realizations for $z$ is within the range $z \in [0.6,1.4]$ in this example.}
    \label{fig:zCDF_example}
\end{figure*}

The velocity dispersion $\sigma$ of a galaxy determines the matter profile according to eq.~\eqref{eq:rhoIstohmlSph} and thus also the solution for the scalar field $\psi$. Since $\sigma$ is only known with a finite accuracy, it results in an uncertainty in $\psi$ and consequently an uncertainty in $\relG$ for the Cepheids in the galaxy. The values for $\sigma$ are listed in Tab.~\ref{tab:sigmalist}. As an example, for N4258 the velocity dispersion is $\sigma = (442 \pm 5) \, \mathrm{km/s}$. With the higher end-value (here, $447 \, \mathrm{km/s}$), the density is greater, hence being more screened (i.e., smaller $\relG$) and vice versa for the lower-end value. To model this uncertainty, we introduce the stochastic variable $z$ which is the value of $\relG$ for a given Cepheid normalized by the value assumed when $\sigma = \sigma_\mathrm{mean}$, that is when the velocity dispersion equals the tabulated mean value,
\begin{equation}
    z = \frac{\relG}{\relG|_{\sigma = \sigma_\mathrm{mean}}}.
\end{equation}
The PDF for $z$ is assumed to be a split normal distribution,
\begin{equation}
    \mathrm{PDF}(z) = \sqrt{\frac{2}{\pi}} \frac{1}{\zmax - \zmin} \left\lbrace \begin{array}{ll}
        e^{- \frac{1}{2} \left( \frac{z-1}{\zmin-1}\right)^2}, & z<1 \\
        e^{- \frac{1}{2} \left( \frac{z-1}{\zmax-1}\right)^2}, & z>1 
    \end{array} \right\rbrace .
\end{equation}
Here, $\zmin$ and $\zmax$ are defined by letting $\sigma$ assume its upper and lower limits, respectively. Thus,
\begin{subequations}
    \begin{align}
    \zmin &= \frac{\relG|_{\sigma = \sigma_\mathrm{mean} + \Delta \sigma}}{\relG|_{\sigma = \sigma_\mathrm{mean}}},\\
    \zmax &= \frac{\relG|_{\sigma = \sigma_\mathrm{mean} - \Delta \sigma}}{\relG|_{\sigma = \sigma_\mathrm{mean}}},
\end{align}
\end{subequations}
where $\Delta \sigma$ is the uncertainty in the velocity dispersion. Since the PDF is different for each Cepheid, depending on its position, $\zmin$ and $\zmax$ depend on $r$. Ideally, $\zmin$ and $\zmax$ should be calculated for each Cepheid (depending on its position) but for simplicity we model the $r$-dependence by fitting linear functions for $\zmin$ and $\zmax$ for each galaxy. The fit is made with respect to ten equidistant points in the range of $r$ from the innermost to the outermost Cepheid in each galaxy. See Fig.~\ref{fig:zCDF_example} for an example.

Integrating the PDF, we obtain the CDF,
\begin{align}
    &\mathrm{CDF}(z) = \nonumber\\
    &\frac{1-\zmin}{\zmax - \zmin} \left\lbrace \begin{array}{ll}
        1 - \mathrm{erf}\left[ \frac{z-1}{\sqrt{2}(\zmin-1)}\right], & z<1 \\
        1 - \frac{\zmax-1}{\zmin-1} \mathrm{erf} \left[ \frac{z-1}{\sqrt{2}(\zmax-1)} \right], & z>1 
    \end{array} \right\rbrace,
\end{align}
where ``erf'' is the error function. This can be inverted, giving $z$ as a function of the CDF-value,
\begin{widetext}
    \begin{equation}
\label{eq:zCDF}
    z(\mathrm{CDF}) = \left\lbrace \begin{array}{ll}
        1 + \sqrt{2} (\zmin-1) \, \mathrm{erf}^{-1}\left[ 1 + \mathrm{CDF} \frac{\zmax-\zmin}{\zmin-1} \right], & \mathrm{CDF} < \frac{1-\zmin}{\zmax-\zmin} \\
         1 + \sqrt{2} (\zmax-1) \, \mathrm{erf}^{-1}\left[ \frac{\zmin-1}{\zmax-1} + \mathrm{CDF} \frac{\zmax-\zmin}{\zmax-1} \right], & \mathrm{CDF} > \frac{1-\zmin}{\zmax-\zmin}
    \end{array} \right\rbrace.
\end{equation}
\end{widetext}
Here, $\mathrm{erf}^{-1}$ is the inverse error function. An example plot of $z(\mathrm{CDF})$ is shown in Fig.~\ref{fig:zCDF_example}. To get a random realization of $\relG$ for a given Cepheid, we insert its distance from the center of the galaxy into the linear fitting functions for $\zmin$ and $\zmax$. Subsequently, we sample a random number between 0 and 1 which we assign to the CDF-value. Using eq.~\eqref{eq:zCDF} to calculate $z$ and multiplying by $\relG|_{\sigma = \sigma_\mathrm{mean}}$ (which we calculate for each Cepheid), we finally obtain a random realization of $\relG$ for the Cepheid according to the distribution. Sampling $\relG$ like this, there can be a small fraction of unphysical values, depending on the number of samples. Therefore, we impose the following physical cutoffs: $0 \leq \relG \leq 2g^2$ (cf. eq.~\eqref{eq:relGmax}).

\subsection{Combining CDFs}
In the sections above, we have shown how the uncertainties in the physical distance and the velocity dispersion affect the value of $\relG$. Here, we describe how we account for both effects simultaneously in a Monte Carlo method. First, we sample a random value of $x$ as in Section \ref{sec:PhysDist} from which we obtain,
\begin{equation}
    \frac{\Delta G}{\GN}' = x \left. \frac{\Delta G}{\GN} \right|_{r=R}.
\end{equation}
This is the (sampled) value that $\relG$ would have if the velocity dispersion would equal its mean value $\sigma = \sigma_\mathrm{mean}$. To take the uncertainty in $\sigma$ into account, we sample a value of $z$ as in Section \ref{sec:VelDisp} and obtain,
\begin{equation}
    \frac{\Delta G}{\GN} = z \frac{\Delta G}{\GN}' = x z \left. \frac{\Delta G}{\GN} \right|_{r=R}.
\end{equation}

\begin{figure*}[t]
    \centering
    \includegraphics[width=0.49\linewidth]{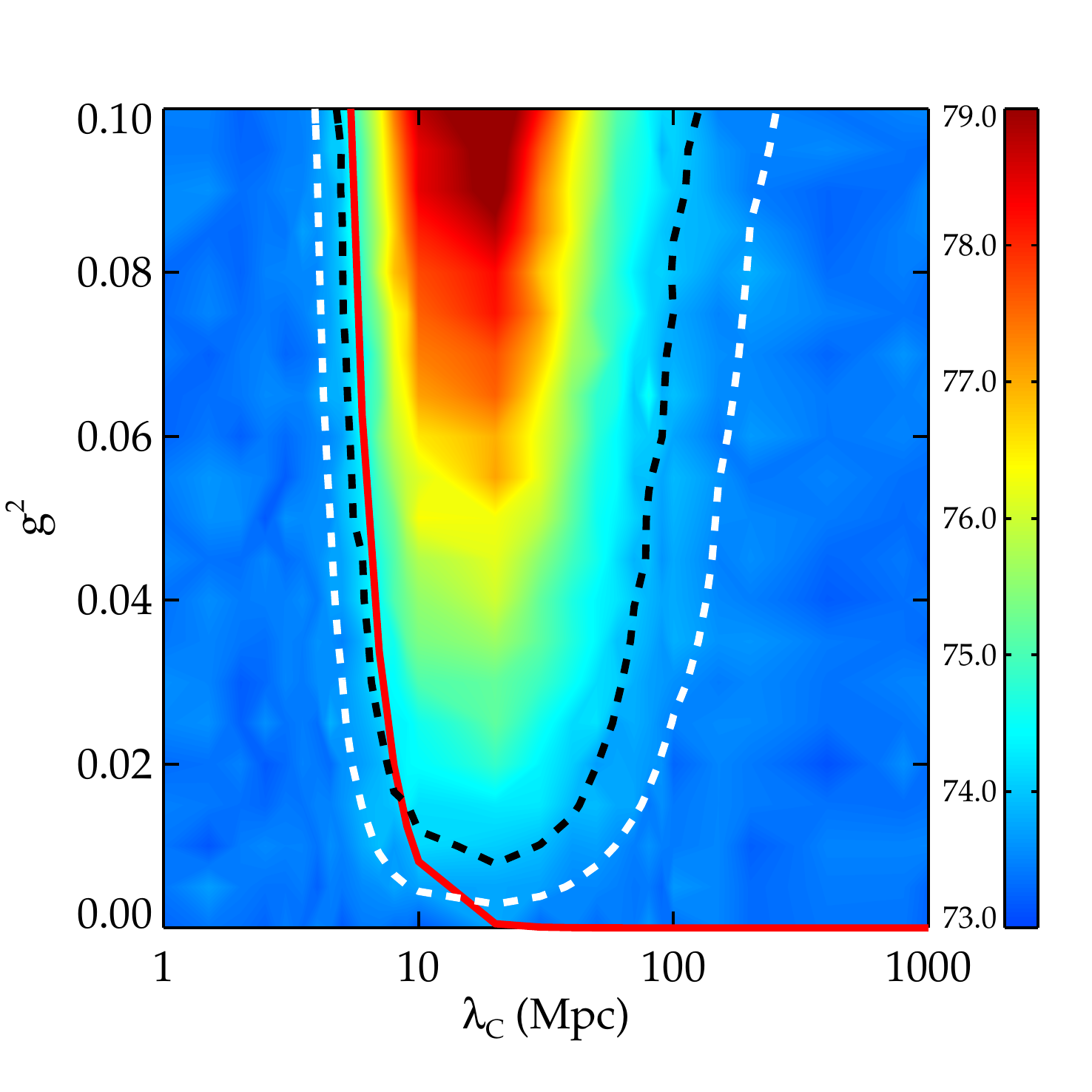}
    \includegraphics[width=0.49\linewidth]{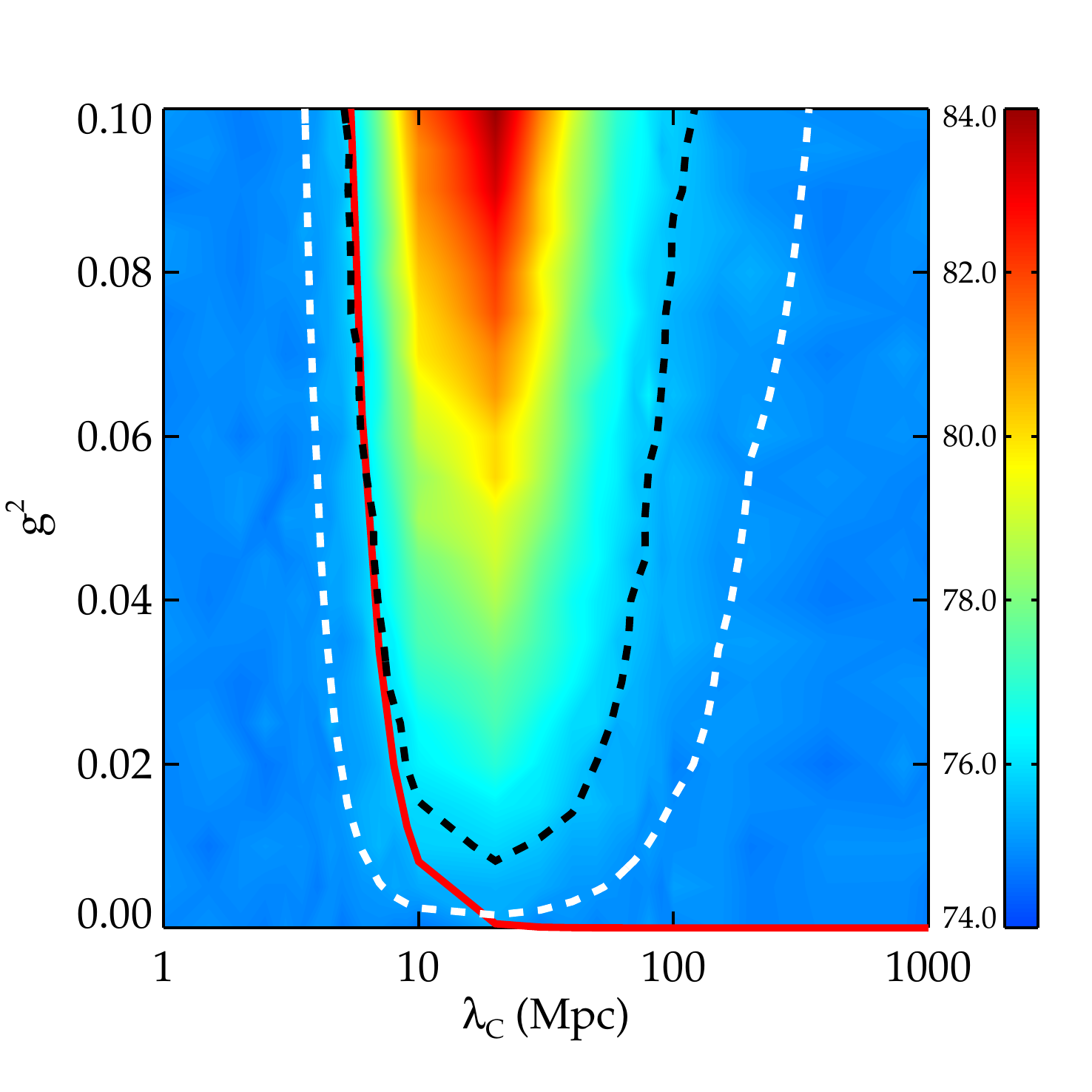}
    \caption{Derived $H_0$ as a function of $\lC$ and $g$ for the symmetron model.
    {\em Left panel:} MW anchor. {\em Right panel:} LMC anchor. The color coding indicates the value of $H_0$. The dashed curves indicate $68.3\%$ confidence contours such that everything above them are excluded by the cosmic distance ladder data to this level. Black curve: constraints from the local Cepheid/SNIa distance ladder. White curve: tension between Cepheid-based and TRGB distances. Red curve: solar system constraints.}
    \label{fig:symk2vslambdaMW}
\end{figure*}

\section{Solar-system tests}
\label{sec:SolarSystTests}
In ref.~\cite{Hinterbichler:2010es}, the authors argue that the MW should be screened for the solar-system tests to be satisfied. In this case, the MW exhibits a thin-shell mechanism, that is $\alpha_\mathrm{MW} \gg 1$. In fact, since $\alpha_\mathrm{MW}$ and $\alpha_{\odot}$ turn out to be of the same order of magnitude, this implies also $\alpha_{\odot} \gg 1$. Thus, the strength of the fifth force compared with the Newtonian force in the solar-system is given by,
\begin{equation}
\label{eq:F5FNrat}
    \frac{F_5}{\FN} = 6 g^2 \frac{\psi_{\mathrm{env},\odot}^2}{\alpha_\odot},
\end{equation}
see eqs.~\eqref{eq:thinshellfactor}-\eqref{eq:relGthinshell}. Now, we want to see how eq.~\eqref{eq:F5FNrat} scales with $\lC$. From eq.~\eqref{eq:alpha}, it follows that $\alpha_\odot \propto 1 / \lC^2$. Further, note that $\psi_{\mathrm{env},\odot} \propto 1/\sqrt{\alpha_\mathrm{MW}}$ (see ref.~\cite{Hinterbichler:2010es} for details) and that $\alpha_\mathrm{MW} \propto 1/\lC^2$. From this, we conclude that,
\begin{equation}
    \frac{F_5}{\FN} \propto g^2 \lC^4.
\end{equation}
Assuming that solar-system tests constrain the fifth force relative to the Newtonian force we can generalize the constraint of ref.~\cite{Hinterbichler:2010es} (i.e., $\lC \lesssim 3 \sqrt{\sa} \, \mathrm{Mpc}$ at $g=1$) to,
\begin{equation}
    g \lesssim \left( \frac{3 \sqrt{\sa} \, \mathrm{Mpc}}{\lC}\right)^2.
\end{equation}

\section{Individual anchors}
\label{sec:IndAnch}
In Fig.~\ref{fig:symk2vslambdaMW}, we show the results when using only the MW as the anchor galaxy (left panel) and using only the LMC as the anchor galaxy (right panel). In the latter case, the inferred value of $H_0$ is similar to when using all anchor galaxies (cf. Fig.~\ref{fig:symk2vslambdaAll}). In the former case with only LMC, $H_0$ assumes larger values, increasing the Hubble tension even further. This is due to LMC being an unusually unscreened galaxy compared with the host galaxies.

\newpage

\bibliography{bibliography}{}
\bibliographystyle{apsrev4-1}

\end{document}